\documentclass[prm,reprint,showpacs,superscriptaddress]{revtex4-1}

\usepackage{hyperref}
\usepackage[utf8]{inputenc}
\usepackage{multirow}
\usepackage{tabulary}
\usepackage{longtable}
\usepackage{siunitx}
\usepackage{threeparttable}
\usepackage{longtable}

\usepackage[T1]{fontenc}
\usepackage[english]{babel}
\usepackage{booktabs}
\usepackage{tikz}
\usepackage{titlesec}
\usepackage{threeparttable}	
\usepackage{appendix}
	\usepackage{MnSymbol}

	\usepackage{hyperref}
	\hypersetup{
		colorlinks,
		citecolor=blue,
		linkcolor=blue,
		urlcolor=blue}

\newcolumntype{K}[1]{>{\centering\arraybackslash}p{#1} }

\begin{document}

\title{Potassium-intercalated bulk HfS$_2$ and HfSe$_2$: Phase stability, structure, and electronic structure}
\author{Carsten Habenicht}
\affiliation{Leibniz Institute for Solid State and Materials Research Dresden, Helmholtzstrasse 20, 01069 Dresden, Germany}
\author{Jochen Simon}
\affiliation{Leibniz Institute for Solid State and Materials Research Dresden, Helmholtzstrasse 20, 01069 Dresden, Germany}
\author{Manuel Richter}
\affiliation{Leibniz Institute for Solid State and Materials Research Dresden, Helmholtzstrasse 20, 01069 Dresden, Germany}
\affiliation{Dresden Center for Computational Material Science (DCMS), TU Dresden, 01062 Dresden, Germany}
\author{Roman Schuster}
\affiliation{Leibniz Institute for Solid State and Materials Research Dresden, Helmholtzstrasse 20, 01069 Dresden, Germany}
\author{Martin Knupfer}
\email{M.Knupfer@ifw-dresden.de}
\affiliation{Leibniz Institute for Solid State and Materials Research Dresden, Helmholtzstrasse 20, 01069 Dresden, Germany}
\author{Bernd Büchner}
\affiliation{Leibniz Institute for Solid State and Materials Research Dresden, Helmholtzstrasse 20, 01069 Dresden, Germany}
\date{\today}

%
%
%
%

\begin{abstract}
We have studied potassium-intercalated bulk HfS$_2$ and HfSe$_2$ by combining transmission electron energy loss spectroscopy, angle-resolved photoemission spectroscopy and density functional theory calculations. The results reveal insights into (1) the intercalation process itself, (2) its effect on the crystal structures, (3) the induced semiconductor-to-metal transitions, and (4) the accompanying appearance of charge carrier plasmons and their dispersions. 

Calculations of the formation energies and the evolution of the energies of the charge carrier plasmons as a function of the potassium content show that certain, low potassium concentrations $x$ are thermodynamically unstable. This leads to the coexistence of undoped and doped domains if the provided amount of the alkali metal is insufficient to saturate the whole crystal with the minimum thermodynamically stable potassium stoichiometry. Beyond this threshold concentration the domains disappear, while the alkali metal and charge carrier concentrations increase continuously upon further addition of potassium. 

At low intercalation levels, electron diffraction patterns indicate a significant degree of disorder in the crystal structure. The initial order in the out-of-plane direction is restored at high $x$ while the crystal layer thicknesses expand by $33-\SI{36}{\%}$. Calculations suggest that this expansion reaches its maximum at doping levels of $x\approx0.25$ before it reverses slightly for higher concentrations. Superstructures emerge parallel to the planes which we attribute to the distribution of the alkali metal rather than structural changes of the host materials. The in-plane lattice parameters change by not more than $\SI{1}{\%}$.

The introduction of potassium causes the formation of charge carrier plasmons whose nature we confirmed by calculating the loss functions and their intraband and interband contributions. The observation of this semiconductor-to-metal transition is supported by calculations of the density of states (DOS) and band structures as well as angle-resolved photoemission spectroscopy.

The calculated DOS hint at the presence of an almost ideal two-dimensional electron gas at the Fermi level for $x<0.6$.

The plasmons exhibit quadratic momentum dispersions which is in agreement with the behavior expected for an ideal electron gas. 
\end{abstract}


\maketitle

\section{INTRODUCTION}
\begin{figure}[b]
	\includegraphics [width=0.46\textwidth]{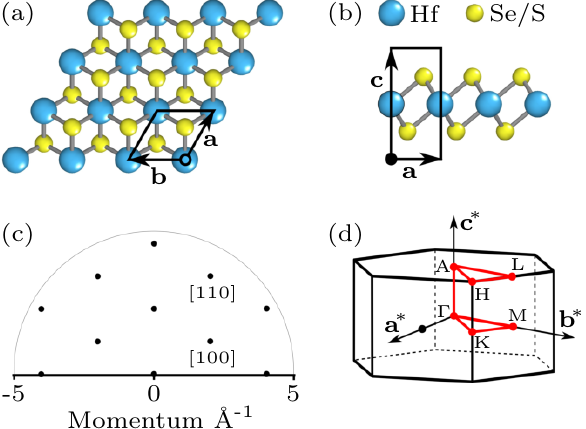}
	\caption{(Color online) (a) Crystal structure and unit cell for the in-plane and (b) out-of-plane directions for pristine HfS$_2$ and HfSe$_2$ with lattice vectors \textbf{a}, \textbf{b} and \textbf{c}. (c) Simulated in-plane electron diffraction pattern and (d) Brillouin zone for those materials with reciprocal lattice vectors $\mathbf{a}^*$ $\mathbf{b}^*$ and $\mathbf{c}^*$ as well as labeled high symmetry points.}
	\label{fig:Crystal}
\end{figure}

Hafnium disulfide and hafnium diselenide are semiconducting transition metal dichalcogenides (TMDCs). Their crystals are formed by slabs (molecular layers) linked by relatively weak Van-der-Waals forces. Each slab consists of an atomic hafnium layer sandwiched between two atomic sulfur/selenium layers [Fig.~\ref{fig:Crystal} (b)]. The compounds typically assume the $1T$-polytype in which a unit cell comprises only one molecular layer and six S/Se atoms are coordinated octahedrally around a hafnium atom (space group: $P\bar{3}m1$, $D^3_{3d}$) resulting in a triangular arrangement of the atoms in the planes [Fig.~\ref{fig:Crystal} (a)] \cite{McTaggart_AJC1958_445,Greenaway_JPCS1965_1445, Conroy_IC1968_459, Bayliss_JoPCSSP1982_1283}. 
The weak interlayer bonding gives rise to quasi-two dimensional properties. This strong anisotropy, band gaps of ${\sim1-\SI{2}{\eV}}$ and predicted electron mobilities as well as sheet current densities \cite{Fiori_NN2014_768} that are significantly higher than in many other TMDCs make the two compounds interesting candidates for electronic devices. So far, transistors \cite{Mleczko_Sa2017_1700481,Kanazawa_SR2016_22277}, field-effect transistors \cite{Kang_APL2015_143108, Kang_N2017_1645,Xu_S2016_3106, Chae_An2016_1309, Chang_JoAP2015_214502, Fu_AM2017_1700439, Nie_AAMI2017_26996, Kaur_NR2018_343, Gong_APL2013_53513}, phototransistors \cite{Xu_AM2015_7881, DeSanctis_NC2018_1652, Yin_APL2016_213105} and photodetectors \cite{Zheng_2M2016_35024, Yan_AFM2017_1702918, Wang_AM2018_1803285, Mattinen_CM2019_5713} have been realized in experiments. The materials are also considered for photovoltaic \cite{Gaiser_PRB2004_75205} and photocatalytic \cite{Singh_CST2016_6605} applications.
\begin{figure}
	\includegraphics [width=0.48\textwidth]{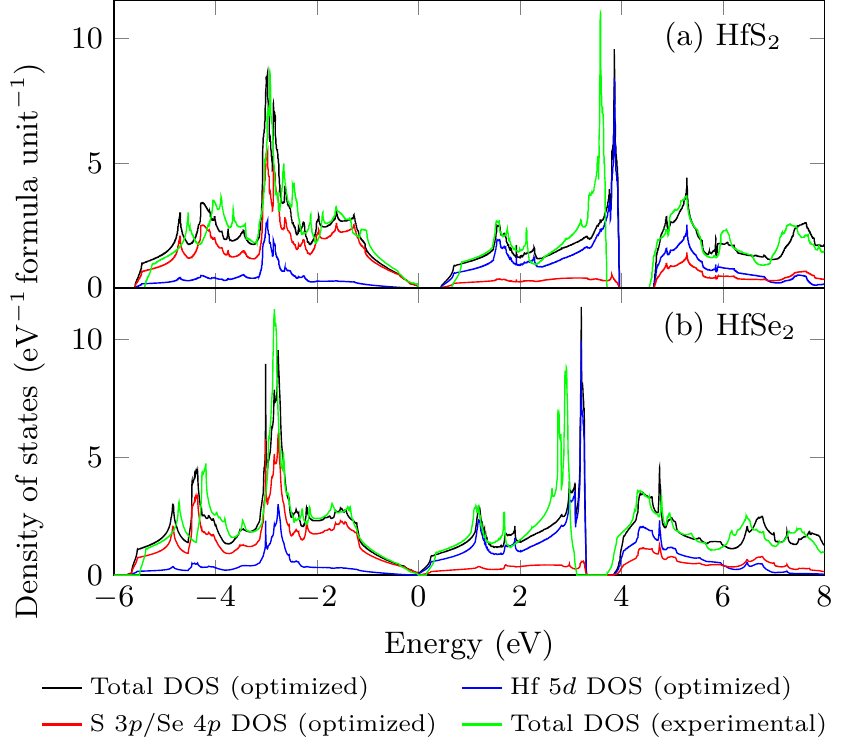}
	\caption{(Color online) Density of states of pristine HfS$_2$ and HfSe$_2$ based on DFT-optimized and experimental lattice parameters.}
	\label{fig:DOS_Undoped}
\end{figure}
Although the available body of research is somewhat smaller compared to other TMDCs such as MoS$_2$, the pristine materials have been investigated with a wide range of experimental techniques such as various optical methods \cite{Yan_AFM2017_1702918, Greenaway_JPCS1965_1445, Terashima_SSC1987_315, Mattheiss_PhysicalReviewB_1973_8_8_3719, Beal_JoPCSSP1972_3531, Lucovsky_PRB1973_3859, Hughes_JoPCSSP1977_1079, Bayliss_JoPCSSP1982_1283, Gaiser_PRB2004_75205, Fong_PRB1976_5442, Borghesi_INCD1984_141, Borghesi_PRB1984_3167, Borghesi_PRB1986_2422, Fu_AM2017_1700439, Yan_AFM2017_1702918}, photoemission \cite{Shepherd_JoPCSSP1974_4416, Jakovidis_JoESaRP1987_275, Kreis_Ass2000_17, Traving_PRB2001_35107, Aretouli_APL2015_143105, Zheng_2M2016_35024, Fu_AM2017_1700439}, x-ray diffraction \cite{Hodul_JoSSC1984_438, Wang_AM2018_1803285}, and Raman spectroscopy \cite{Iwasaki_JotPSoJ1982_2233, Kanazawa_SR2016_22277, Xu_AM2015_7881, Chae_An2016_1309, Zheng_2M2016_35024, Ibanez_SR2018_12757, Fu_AM2017_1700439, Nie_AAMI2017_26996, Yan_AFM2017_1702918, Kaur_NR2018_343, Najmaei_S2018_1703808, Mattinen_CM2019_5713, Roubi_PRB1988_6808, Kang_APL2015_143108, Yin_APL2016_213105, Kang_N2017_1645, Cruz_MC2018_1191, Cingolani_PS1988_389, Wang_AM2018_1803285}.
But also resistivity \cite{Hodul_JoSSC1984_438, Zheng_JotPSoJ1989_622, McTaggart1958_471, Radhakrishnan_AJC2008_283}, conductivity \cite{Conroy_IC1968_459,Najmaei_S2018_1703808} Hall coefficient,\cite{Zheng_JotPSoJ1989_622}, magnetic susceptibility \cite{Conroy_IC1968_459}, scanning transmission electron microscopy \cite{Aretouli_APL2015_143105}, and electron energy-loss \cite{Bell_AdvancesinPhysics_1976_25_1_53, Habenicht_PRB2018_155204} experiments were performed. The investigations were supported by numerous theoretical approaches \cite{Murray_JoPCSSP1972_746, Mattheiss_PhysicalReviewB_1973_8_8_3719, Fong_PRB1976_5442,Bullett_JoPCSSP1978_4501, Traving_PRB2001_35107, Zhang_NR2014_1731, Lebegue_PRX2013_31002, Eknapakul_PRB2018_201104, Chae_An2016_1309, Nie_AAMI2017_26996, Reshak_PBCM2005_25, Zhao_pssb2017_1700033, Jaiswal_SST2018_124014, Najmaei_S2018_1703808}.

The weak Van-der-Waals interactions among the molecular layers allow introducing intercalants into the gaps between the slabs. This could alter the properties of the compounds significantly by affecting the atomic arrangements, band structures and band fillings, depending on the intercalants' sizes and electro-negativities. This variability of characteristics makes intercalated TMDCs interesting for new applications and as research objects to obtain a better understanding of fundamental physical phenomena. Besides transition metals \cite{Yacobi_JoPCSSP1979_2189, Iwasaki_SM1983_157, Pleshchev_PotSS2011_2054}, which have been intercalated into both materials, alkali metals have been used as electron donors for HfSe$_2$ \cite{Dines_MRB1975_287, Whittingham_Mrb1975_363}. Lithium \cite{Onuki_JotPSoJ1982_880, Beal_PMB1981_965} and sodium \cite{Mleczko_Sa2017_1700481, Eknapakul_PRB2018_201104} doping leads to a transition from semiconducting to metallic behavior.
In the pristine materials, the valence states are mainly comprised of S/Se $p$-orbitals which are almost filled due to a transfer of electrons from the hafnium atoms \cite{Camassel_INCB11977_185}.

The conduction bands are formed largely by empty Hf $d$-states \cite{McTaggart1958_471, Camassel_INCB11977_185} that are split into $e_{g}$ 
and $t_{2g}$ 
orbitals \cite{Gong_APL2013_53513}. The nature of the orbitals comprising the valence and conduction bands is reflected in the calculated density of states (DOS) in Fig. \ref{fig:DOS_Undoped}. Intercalated alkali-metal atoms donate electrons to the lowest conduction bands of the host materials leading to the mentioned semiconductor-to-metal transition.

This transition made the two hafnium compounds attractive for a study using transmission electron energy-loss spectroscopy (EELS) supported by angle-resolved photoemission spectroscopy (ARPES).
In particular, EELS permits the momentum dependent measurement of plasmons, the collective oscillations of charge carriers, typically associated with the free electron gas in metals \cite{Nozieres_PR1959_1254, Raether_2006_}. We combined the two experimental methods with density-functional theory (DFT) calculations to shed light on the effects of doping on crystal structures, the semiconductor-to-metal transitions, the charge carrier plasmons, and the dimensionality of the conduction bands of the title compounds.
This research is a continuation of our previous efforts to investigate the effects of alkali-metal doping on TMDCs via EELS for K$_x$TaS$_2$, Na$_x$TaSe$_2$, K$_x$NbSe$_2$, Na$_x$NbSe$_2$ \cite{Mueller_PRB2016_35110}, K$_x$TaSe$_2$ \cite{Koenig_EL2012_27002, Koenig_PRB2013_195119}, K$_2$WSe$_2$ \cite{Ahmad_JoPCM2017_165502} and K$_x$MoS$_2$ \cite{Habenicht_Sfp_}.
%
%

\section{EXPERIMENT}\label{sec:Experiment}
\subsection{EELS Experiments}
EELS is a bulk sensitive scattering technique. The spectra are proportional to the loss function $L(\mathbf{q},\omega)=\text{Im}[-1/\epsilon (\mathbf{q},\omega)]$ \cite{Sturm_ZNA1993_233}. In this equation, $\epsilon (\bf{q},\omega)$ is the energy \footnote{Please note that the terms energy and frequency are used synonymously in the work ($E=\hbar\omega$).} $\omega$ and momentum \textbf{q} dependent dielectric function. For our experiments, we purchased bulk single crystals of hafnium disulfide and hafnium diselenide from HQ Graphene and cleaved them \textit{ex situ} into thin films of approximately $\SI{100}{\nano\meter}$ thickness using adhesive tape. Mounted on platinum transmission electron microscopy grids, the samples were measured in a purpose-built transmission electron energy-loss spectrometer operating with a primary electron energy of $\SI{172}{\kilo\eV}$ and equipped with a helium flow cryostat (see Refs. \onlinecite{Fink_AEEP_1989_75__121, Roth_JournalofElectronSpectroscopyandRelatedPhenomena_2014_195__85} for more detailed descriptions of the instrument).

The intercalation was performed by thermally evaporating potassium from SAES alkali metal dispensers onto the samples in an ultrahigh vacuum chamber (base pressure below $\SI[parse-numbers=false]{10^{-10}}{\milli\bar}$) directly attached to the instrument. The films were placed in the potassium vapor for time periods ranging from 15 to $\SI{90}{\second}$ and subsequently annealed for approximately $1-\SI{2}{\hour}$ at $80-\SI{200}{\celsius}$ between each EELS measurement to attain a series of increasing doping levels in the same samples. The highest achieved potassium concentrations were 0.90 and 1.25 for HfS$_2$ and HfSe$_2$, respectively (see Sec.~\ref{sec:ExpIntercalationLevel} for a description of the method used to calculate the doping levels) because additional intercalation attempts did not change the diffraction patterns nor the loss spectra.

For an energy range between 0.2 and $\SI{70}{\eV}$, the EELS spectra for the pristine and intercalated materials were acquired for a momentum transfer of $|\mathbf{q}|=\SI{0.1}{\per\angstrom}$ in the $\Gamma K$ and $\Gamma M$ directions of the Brillouin zone (see Fig. \ref{fig:Crystal} (d) for a picture of the Brillouin zone and its high symmetry points).
The same was done for spectra up to $\SI{10}{\eV}$ for various |\textbf{q}| between 0.075 and $\SI{0.7}{\per\angstrom}$. 
We also obtained electron diffraction patterns along those orientations in the momentum region $\SI{0.2}{\per\angstrom}\!<|\mathbf{q}|<\SI{5.0}{\per\angstrom}$ at zero energy transfer ($E=\SI{0}{\eV}$). For selected doping levels, sets of such diffraction patterns were measured consecutively in $\SI{1}{\degree}$ steps across at least one half of the Brillouin zone and subsequently combined to form in-plane diffraction maps. In all those cases, the energy and momentum resolutions were $\Delta E=\SI{82}{\milli\eV}$ and $\Delta |\mathbf{q}|=\SI{0.04}{\per\angstrom}$.
Moreover, core level spectra of sulfur/selenium and potassium were obtained with resolutions of $\Delta E=\SI{369}{\milli\eV}$ and $\Delta |\mathbf{q}|=\SI{0.07}{\per\angstrom}$. 

\begin{figure}[b]
	\includegraphics [width=0.48\textwidth]{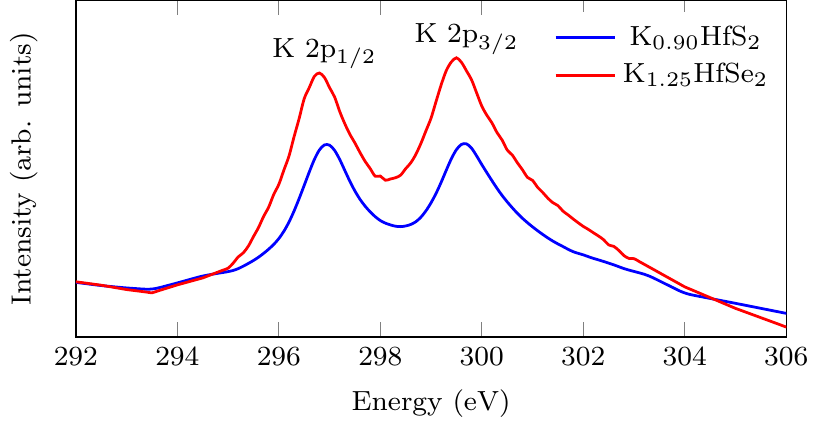}
	\caption{(Color online) EELS spectra of the K $2p$ core levels for the highest achieved intercalation levels.}
	\label{fig:CoreLevels}
\end{figure}

Long-time EELS measurements did not reveal noticeable beam damage in the crystals. 
Decomposition effects such as the formation of salts (e.g. K$_2$S or K$_2$Se), which were reported for MoS$_2$ highly doped with potassium \cite{Somoano_TheJournalofChemicalPhysics_1973_58_2_697, Zhang_Nl2015_629}, sodium \cite{Wang_AN2014_11394} or lithium \cite{Cheng_An2014_11447, Huang_SCC2018_222}, were not observed in this investigation. Such a chemical reaction would cause a splitting of the two K $2p$ core level peaks in the EELS spectra because of the concurrent presence of 
potassium in the salt and potassium in the Van-der-Waals gaps. The K $2p$ spectra in Fig. \ref{fig:CoreLevels} do not show such a behavior even at the highest achieved alkali metal concentrations.

The $ab$-planes of the two materials align with the sample surfaces and are initially positioned perpendicular to the electron beam. In this configuration, the crystal planes are parallel to the momentum transfer of the scattered electrons which lies in a plane perpendicular to the beam. The instrument allows to rotate the sample surfaces up to $\SI{45}{\degree}$ with respect to the momentum transfer plane. We will refer to this angle as polar angle which is $\SI{90}{\degree}$ offset from the angle of incidence. The spectrometer does not permit diffraction measurements directly in the $c$-direction. However, information for this direction can be acquired by adjusting the polar angle until the reciprocal lattice points of an adjacent crystal lattice layer are aligned with the momentum transfer plane such that the diffraction peak associated with the neighboring plane can be detected. The momentum positions of two Bragg peaks that are equivalent in the planes but not in the out-of-plane direction (e.g. [110] and [111]) can than be related via the Pythagorean theorem to calculate the separation of the planes in momentum space and the layer thickness in real space. This approach may be repeated for successive planes (e.g. [110], [111], [112],...) up to the maximum polar angle of $\SI{45}{\degree}$.

In an effort to deduce the unscreened plasmon frequencies from the spectra of the intercalated samples, Kramers-Kronig analyses were carried out. 
The resulting optical conductivity functions were fitted according to the Drude-Lorentz model. This classical approach models the dielectric function $\epsilon(\omega)$ as a function of the frequency $\omega$ of a series of oscillators representing the excitation of the involved free (Drude term) and bound (Lorentz term) charges \cite{Li_2017_,Hecht_Optics_2017}:
\begin{equation}
\epsilon(\omega)=\epsilon_{\infty}-\underbrace{\frac{\omega_{p}^2}{\omega^2+i\gamma\omega}}_\text{Drude term}+\underbrace{
\sum_{j=1}\frac{\omega_{pj}^2}{\omega_j^2-\omega^2-i\gamma_j\omega}}_\text{Lorentz terms}.\label{equ:epsilon}\\
\end{equation}
Here, $j$ is the index number of each oscillator in the sum, $\omega_j$ refers to the resonant frequency of the $j$-th oscillator, $\gamma$ and $\gamma_j$ represent the frequency widths (damping factors) of the Drude and $j$-th Lorentz oscillator, respectively. Further, $\epsilon_{\infty}$ is the background dielectric constant combining the effects of oscillators not included in the sum.

The oscillator strengths are expressed by the plasmon frequencies $\omega_{p}$ ($\omega_{pj}$):
\begin{subequations}
\label{equ:PlasmonFrequ}
\begin{flalign}
	\text{for the Drude term:\;}\omega_{p}=\sqrt{\frac{q_0^2 n}{m^*_{e}\epsilon_0}}\;\text{and} \label{equ:PlasmonFrequDrude}\\
	\text{for the Lorentz terms:\;}\omega_{pj}=\sqrt{\frac{q_0^2 n_j}{m^*_{ej}\epsilon_0}.}\label{equ:PlasmonFrequLorentz}
\end{flalign}
\end{subequations}
Here, $\epsilon_0$ is the permittivity of free space, $q_0$ the elementary electron charge, $n$ ($n_j$) the electron density and $m^*_e$ ($m^*_{ej}$) the effective electron mass related to the Drude oscillator (the $j$-th Lorentz oscillator).
As defined in Equ. \ref{equ:PlasmonFrequ}a, $\omega_p$  represents the unscreened plasmon frequency of the free electrons. It differs from the screened plasmon frequency reflected in electron energy-loss spectra due to damping by single particle excitations in the surrounding host material.

\subsection{Calculation of Intercalation Levels in EELS Measurements}\label{sec:ExpIntercalationLevel}
Due to the setup of the experiments, it was not possible to measure the potassium concentration $x$ in the crystals in a direct way. As an alternative, the charge carrier plasmon peak position (see Sec.~\ref{sec:STM_Transition}) was extracted from the EELS response associated with the intercalation step that produced the most pronounced plasmon peak in an intercalation series. The peak position was determined after eliminating the effects of the quasielastic line by fitting the latter with a Gaussian function and the plasmon feature with the loss function of a Drude oscillator (see Ref.~\onlinecite{Schuster_PRB2009_45134} for details). The peak energies found in this way were matched with the interpolated plasmon peak energies from the DFT loss spectra (see Sec.~\ref{sec:STM_Transition}) for compounds with various simulated potassium stoichiometries. This comparison of experimental and simulated plasmon peak energies allowed the assignment of the
interpolated simulated doping levels to the experimental intercalation step. The specific spectra to which the described procedures were applied turned out to be the ones with doping levels of $x=0.55$ for HfS$_2$ and $x=0.70$ for HfSe$_2$.
All other doping levels were determined by calculating the areas under the K $2p$ core level peaks for each intercalation step. The integration was performed after deducting a linear background between 293.5 and $\SI{308}{eV}$ from the spectra. The fractional changes of each area relative to the area for which the doping levels were determined from the plasmon peak positions were multiplied with the concentrations stated above ($x_{\mathrm{HfS_2}}=0.55$ and $x_{\mathrm{HfSe_2}}=0.70$) to find the potassium concentrations for the other intercalation steps. 

\subsection{Photoemission Measurements}\label{sec:ExpARPES}
The photoemission measurements were performed at room temperature in an instrument with a base pressure of $\sim\SI{e-10}{\milli\bar}$ equipped with a Scienta R4000 electron analyzer, a helium discharge lamp with a photon energy of $\SI{40.81}{\eV}$ (He II) and an Al $K_{\alpha}$ x-ray source with energy of $\SI{1486.6}{\eV}$. The crystals were attached to copper sample holders with electrically conductive EPO-TEK H27D epoxy and cleaved \textit{in situ}. The potassium deposition on the samples was achieved by thermal evaporation from SAES alkali metal dispensers. The Fermi energy E$_f$ was determined by fitting the Fermi edge of a gold sample measured under the same conditions. The alkali metal concentrations in the samples were derived from the fractions of the cross section adjusted Hf $4f$ 
and K $2p$ XPS core level peak areas.

The experimental setup does not allow an \textit{in situ} transfer of the samples between the EELS and the ARPES spectrometers. Consequently, it was not possible to apply both methods to the same specimens and intercalations had to be performed in each instrument separately.

\section{COMPUTATIONAL DETAILS}\label{sec:ComputationalDetails}
To support the interpretation of the experimental results, we carried out density function theory calculations using the 18.00-52 version\footnote{\lowercase{https://www.fplo.de/}} of the full-potential local-orbital code (FPLO) \cite{Koepernik_PRB1999_1743, Richter_2008_271}. The Perdew-Wang-92 exchange-correlation-functional \cite{Perdew_PRB1992_13244} of the local density approximation (LDA) was employed. The linear tetrahedron method with Bl\"ochl corrections was applied for $k$-space integrations.
We checked the importance of the spin-orbit interaction for the case of HfSe$_2$. A comparison of the DOS of HfSe$_2$ with and without spin-orbit interaction showed differences of up to $\sim \SI{100}{\milli \eV}$ in the band edges but an almost perfect overall agreement in the whole relevant energy range between $-\SI{5}{\eV}$ and $+\SI{10}{\eV}$ (not shown). Hence, we decided to perform all further calculations only in the scalar relativistic mode.

Calculations were carried out for undoped bulk compounds and for certain structural configurations of K-doped bulk materials. The pristine crystals consist of planes in the sequence: A (Hf) - B (S/Se) - C (S/Se) - A(Hf). The intercalated bulk structures for potassium concentrations up to $x=1$ were constructed by placing K-atoms into the Van-der-Waals gaps such that they occupy the A positions in the sequence: A (Hf) - B (S/Se) - A(K) - C (S/Se) - A(Hf) (see Fig. \ref{fig:Supercells} (a) in Appendix \ref{sec:AppendixA}). Supercells extending in the $ab$-plane were set up containing the appropriate number of potassium atoms to achieve the desired stoichiometries (see Fig. \ref{fig:Supercells} (b) - (j) in Appendix \ref{sec:AppendixA}). 

For all considered structures, an iterative process was used to optimize the lattice parameters. In each iteration step, one of the lattice constant was changed $\SI{0.01}{\angstrom}$ and the atomic positions were optimized by minimizing the total energy within the limitations of the applied space groups and an accuracy of $\SI{1e-3}{\eV\per\angstrom}$ on each atom. 

In addition, we also used experimental lattice constants for the undoped materials. In those cases, only the internal parameters were relaxed. The calculated lattice parameters underestimated the experimental values by $\SI{2}{\%}$ or less which is typical for LDA calculations \cite{Jiang_TJoCP2011_204705}.

We should note that, materials with important contributions of Van-der-Waals bonding require the consideration of the related dispersion forces for accurate structure optimization. Here, we neglect these contributions and instead rely on error compensation with the known LDA overbinding. The obtained realistic lattice parameters $c$ (Fig. \ref{fig:CalcParameters}) provide an {\it a posteriori} justification for our approach.

All used space groups, $k$-meshes, optimized atomic coordinates as well as experimental and optimized lattice parameters are listed in Tables \ref{tab:HfS2_CalcDetails} and \ref{tab:HfSe2_CalcDetails} of Appendix \ref{sec:AppendixA}. The band structures, densities of states, formation energies, and optical properties, in particular the loss functions and their interband and intraband contributions, were calculated for all structures described in those tables. The intraband contributions correspond to the Drude terms in Equ. \ref{equ:PlasmonFrequDrude} modeling the behavior of the free electrons in a metal and, therefore, reflect charge carrier plasmons. They are collective oscillations of all conduction electrons. Frequency widths of $\gamma=\SI{0.5}{\eV}$ and $\SI{0.3}{\eV}$ were applied to the Drude contributions (Equ. \ref{equ:epsilon}) of HfS$_2$ and HfSe$_2$, respectively, to match the calculated plasmon peak widths to the experimental spectra.

In order to test convergence, calculations with finer $k$-meshes ($104\times104\times56$) were performed for the pristine bulk materials with optimized lattice constants but did not yield any relevant improvements in the DOS. 

ARPES measurements are surface sensitive since the contributions of the photoelectrons to the intensity $I$ of the spectra decrease with the distance $d$ of the emitting atoms from the sample surface: 
\begin{equation}
I=I_0e^{(-d/d_0)}.
\label{equ:IntDecay}
\end{equation}
where $d_0$ denotes the characteristic escape depth. 
To simulate the band structures produced by the ARPES measurements, we constructed supercells consisting of 5 molecular crystal layers separated by $\SI{20}{\angstrom}$ of vacuum while retaining the optimized unit cell parameters, bulk atomic spacing and bond angles of the bulk layers. 
The computational and structural details are also given in Tables \ref{tab:HfS2_CalcDetails} and \ref{tab:HfSe2_CalcDetails} of Appendix \ref{sec:AppendixA}. For the presentation of the band structures, the contributions of the individual atomic layers were weighted based on their distance from the crystal-vacuum border according to Equ. \ref{equ:IntDecay} to account for the decay in intensity. The escape depth for a photon energy of $\SI{40.81}{\eV}$ was assumed to be $d_0=\SI{4}{\angstrom}$ based on the compilation of inelastic mean free path measurements (universal curve) published by Seah and Dench \cite{Seah_SaIA1979_2}.

It should be pointed out that we did not attempt to mimic the exact doped crystal structures observed in the experimental diffraction patterns as this would have been too extensive given the potential effects of disorder and large wavelength modulations. Nevertheless, the generated results turn out to be realistic enough to provide meaningful support for our interpretation of the experimental findings.

\section{Results and Discussion}\label{sec:R&D}
\subsection{Diffraction Patterns}\label{sec:DiffractionPatterns}
\begin{figure*}
	\includegraphics [width=\textwidth]{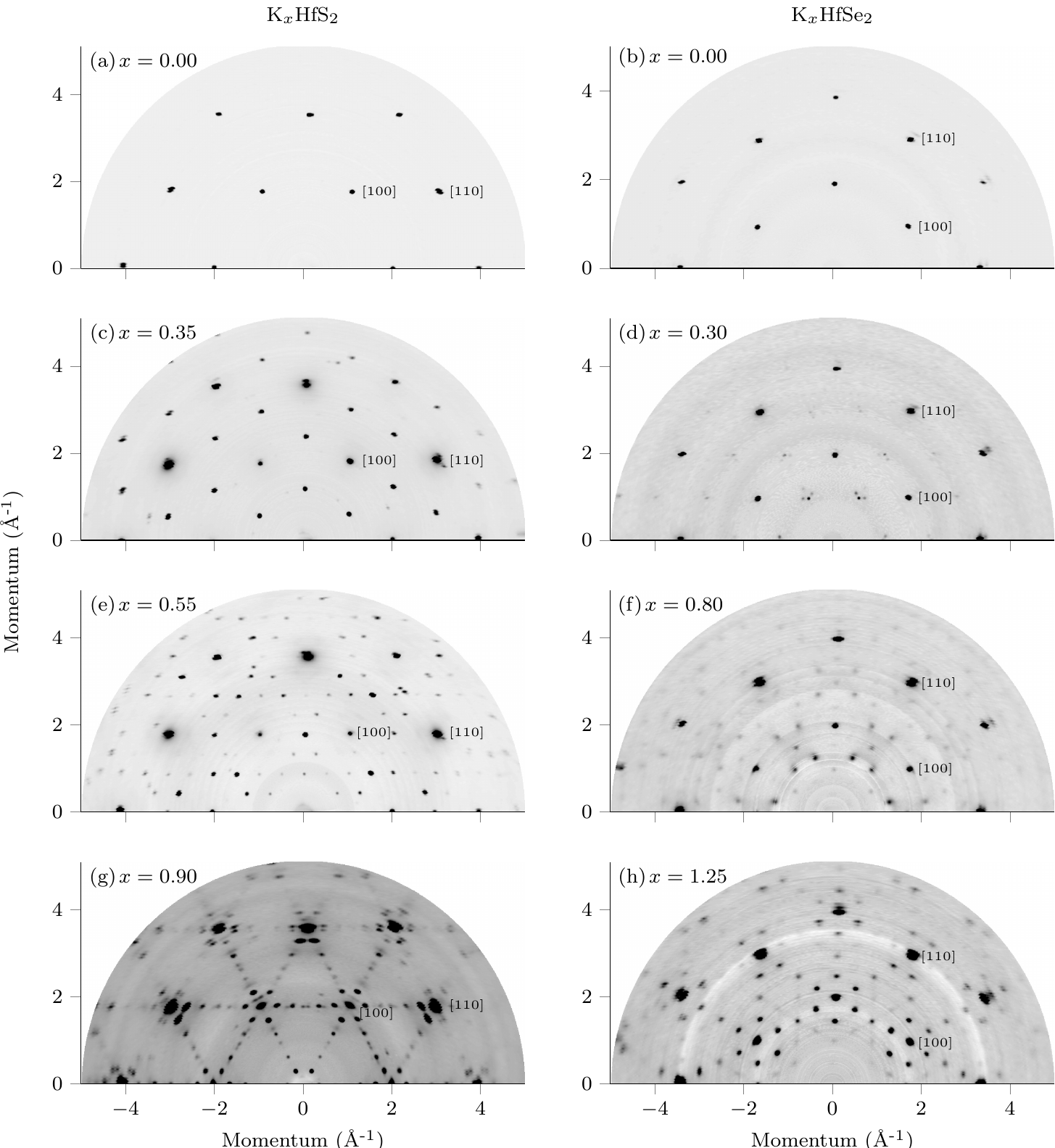}
	\caption{(Color online) Diffraction patterns in the $ab$-plane of pristine and K-intercalated HfS$_2$ (left column) and HfSe$_2$ (right column). All data were measured using elastic electron scattering at $T=\SI{20}{\K}$. Without a particular reason, the maps for K$_x$HfS$_2$ and K$_x$HfSe$_2$ were acquired in such a way that they are rotated by $\SI{30}{\degree}$ with respect to each other.}
	\label{fig:Diffr_Map}
\end{figure*}

\begin{figure*}
	\includegraphics [width=\textwidth]{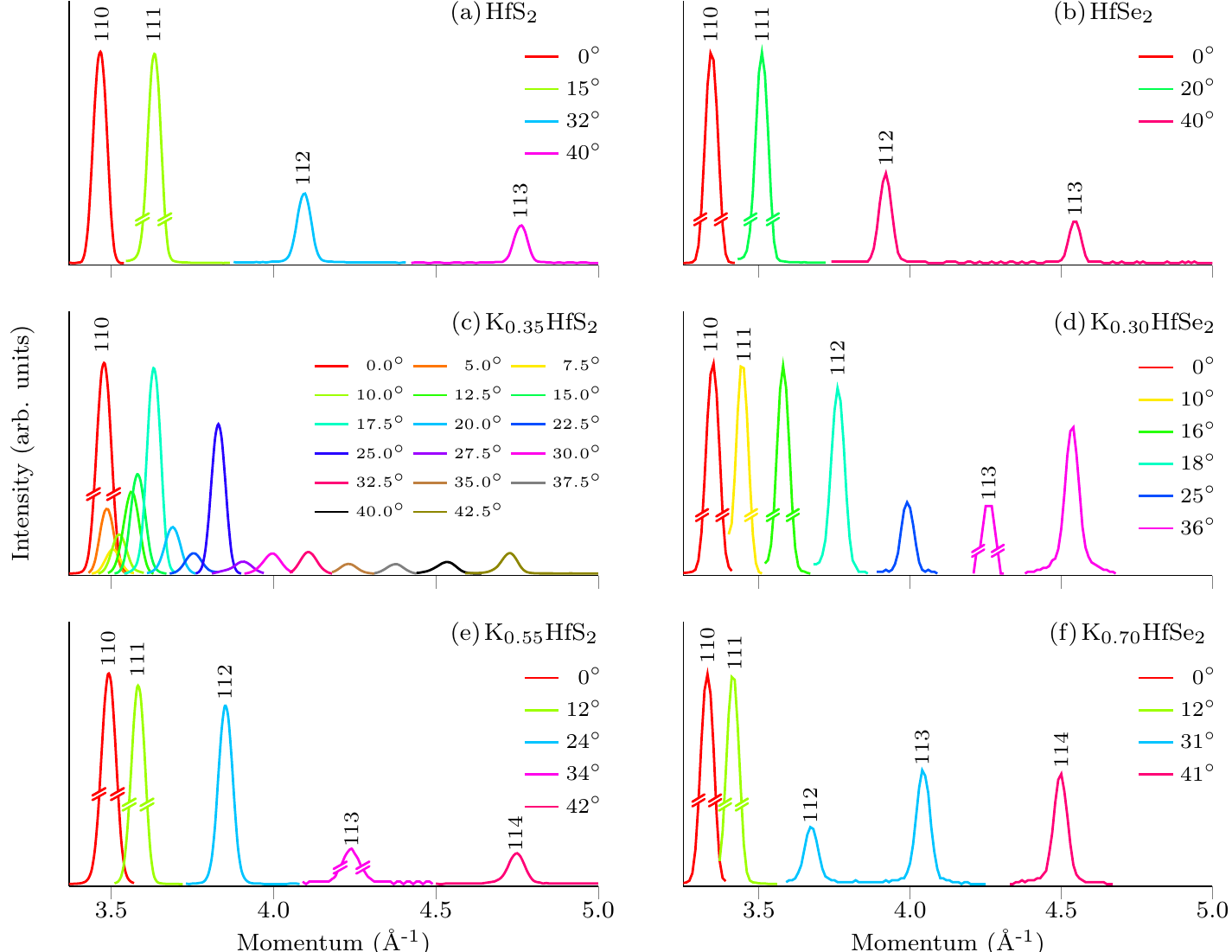}
	\caption{(Color online) Diffraction peaks for the first [11$w$] reflexes of pristine and K-intercalated HfS$_2$ (left column) and HfSe$_2$ (right column). The angles labeling the curves refer to the polar angles used to rotate the lattice planes associated with the respective $w$-values in the spectrometer's plane of momentum transfer. All data were measured using elastic electron scattering at $T=\SI{20}{\K}$.}
	\label{fig:Diffr_cAxis}
\end{figure*}

Electron diffraction patterns were acquired to observe the effect of the potassium intercalation on the crystal structure. For pure HfS$_2$ and HfSe$_2$, Fig. \ref{fig:Diffr_Map} (a) and (b) show the expected hexagonally arranged in-plane diffraction spots [see Fig. \ref{fig:Crystal} (c) for the simulated diffraction pattern]. The diffraction peaks for a number of [11$w$] reciprocal lattice points are presented in Fig. \ref{fig:Diffr_cAxis} (a) and (b). We were able to index the peaks unambiguously confirming that the crystals are of the $1T$ polytype and of high crystallinity. The Bragg peak locations translate to real space lattice constants of $a=\SI{3.63}{\angstrom}$ and $c=\SI{5.77}{\angstrom}$ for the sulfide compound as well as $a=\SI{3.76}{\angstrom}$ and $c=\SI{6.06}{\angstrom}$ for the selenide compound (see Table \ref{tab:DiffrPP} in Appendix \ref{sec:AppendixA} for the peak index and position details in the $c$-direction). The in-plane parameters, measured at $T=\SI{20}{\K}$ agree very well with and the $c$-parameters are only less than $\SI{1.5}{\%}$ lower than values published by others (mostly measured at room temperature) \cite{Lucovsky_PRB1973_3859, Hodul_JoSSC1984_438, Greenaway_JPCS1965_1445, Friend_AiP1987_1, McTaggart_AJC1958_445, Conroy_IC1968_459, Rimmington1974, Zheng_JotPSoJ1989_622, Whittingham_Mrb1975_363}.
%
%
\begin{table}
\setlength\extrarowheight{1.3pt}
\centering
\renewcommand{\tabcolsep}{0.12cm}
\caption{Lattice expansions in $c$-direction upon potassium intercalation.}
\label{tab:cLatticeExp}
\begin{tabular}{ccccccccc}
\toprule
\toprule
\multicolumn{4}{c}{HfS$_2$}	&&\multicolumn{4}{c}{HfSe$_2$}	\\\cline{1-4} \cline{6-9}
Doping	&$c$-para-						&\multicolumn{2}{c}{Expan-}	&&Doping	&$c$-para-	&\multicolumn{2}{c}{Expan-}	\\ 
level	&meter						&\multicolumn{2}{c}{sion}	&&level		&meter	&\multicolumn{2}{c}{sion}	\\ \cline{3-4} \cline{8-9}
x		&$\left(\SI{}{\angstrom}\right)$		&$\left(\SI{}{\angstrom}\right)$	&\%&&x			&$\left(\SI{}{\angstrom}\right)$	&$\left(\SI{}{\angstrom}\right)$	&\%	\\
\midrule
0.00	&5.77	&-		&-		&&0.00	&6.06	&-&-	\\
0.55	&7.84	&2.07	&35.9	&&0.70	&8.23	&2.17	&35.8	\\
0.60	&7.80	&2.03	&35.3	&&0.80	&8.09	&2.03	&33.4	\\
\bottomrule
\bottomrule
\end{tabular}
\end{table}
\begin{figure}[b]
	\includegraphics [width=0.47\textwidth]{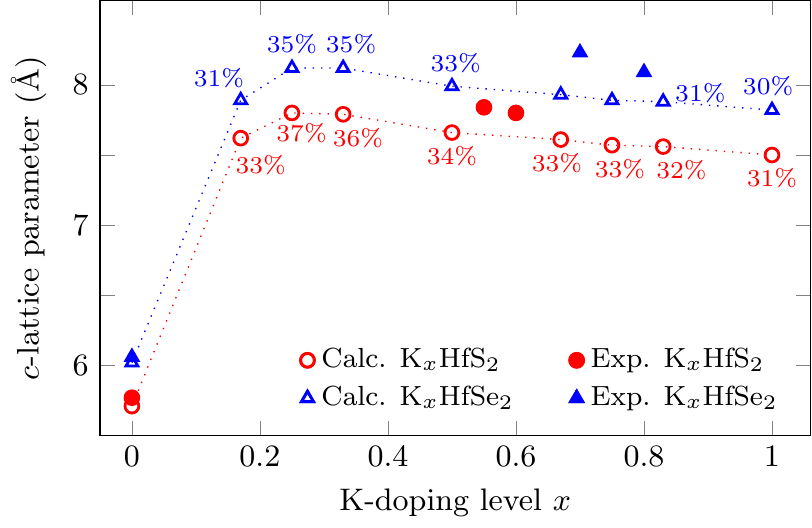}
	\caption{(Color online) Calculated and experimental $c$-lattice parameters as a function of doping level. The percentage values next to the data points indicate the lattice expansions resulting from those data.}
	\label{fig:CalcParameters}
\end{figure}

At a doping level of 0.35, HfS$_2$ exhibits a clear $\sqrt{3}a\times\sqrt{3}a$ superstructure [Fig. \ref{fig:Diffr_Map} (c)] where $a$ refers to the lattice parameter for the undoped material in the $ab$-plane. This is similar to what has been observed in alkali metal doped MoS$_2$ where this diffraction pattern is caused by a distorted host lattice referred to as $1T'''$ structure \cite{Habenicht_Sfp_, Fang_JotACS2019_790}. It also fits to the $x = 0.33$ superstructure used for our DFT calculations, see Fig. \ref{fig:Supercells} (e).

However, we were not able to index the Bragg peaks associated with the [11$w$] reciprocal lattice points for that potassium concentration [Fig. \ref{fig:Diffr_cAxis} (c)]. This indicates a high degree of disorder along the $c$-direction. It appears that the unit cell is not uniformly limited to only one molecular layer. Consequently, it is unlikely that the in-plane superstructure is caused by the occurrence of a $1T'''$ crystal structure. Additional potassium ($x=0.55$) lead to a significant reduction in the intensity of the $\sqrt{3}a\times\sqrt{3}a$ pattern and the appearance of additional diffraction spots in the $ab$-plane [Fig. \ref{fig:Diffr_Map} (e)]. The measurements of the [11$w$] diffraction spots [Fig. \ref{fig:Diffr_cAxis} (e)], however, can be indexed which shows a restored order along the $c$-axis with one layer per unit cell which is retained for all higher doping levels. This finding justifies the use of the same, smallest possible, periodicity in $c$-direction in our DFT calculations.
The reduced peak distances signify a considerable lattice expansion perpendicular to the planes. At the highest achieved alkali metal concentration of $x=0.90$, the in-plane diffraction pattern is rearranged again to show mainly five diffraction peaks the lines between the main host lattice spots [Fig. \ref{fig:Diffr_Map} (g)]. 

The intercalation of HfSe$_2$ resulted in weak multi-peak clusters distributed in a $\sqrt{3}a\times\sqrt{3}a$ manner at $x=0.30$ [Fig. \ref{fig:Diffr_Map} (d)]. At the same time, the crystal becomes inhomogeneous in the $c$-direction [Fig. \ref{fig:Diffr_cAxis} (d)] which permanently reverts back to a $1T$ order at $x=0.70$ with increased lattice parameter $c$ [Fig. \ref{fig:Diffr_cAxis} (f)]. Nevertheless, the in-plane diffraction pattern undergoes further changes [Fig. \ref{fig:Diffr_Map} (f)] until it settles at a $4a \times 4a$ superstructure for $x=1.25$ [Fig. \ref{fig:Diffr_Map} (h)].

Given the almost continuous, doping level-dependent change of the diffraction patterns in the $ab$-plane, it is unlikely that this behavior is caused by structural changes in the host crystals. Such phase changes have been reported for MoS$_2$ where the Fermi level separates the fully occupied $4d^2_z$ orbital from the other $4d$ orbitals leading to a structural instability upon electron doping \cite{Kertesz_JACS1984_3453, Enyashin_TJoPCC2011_24586, Enyashin_CaTC2012_13,Chhowalla_Nc2013_263, Voiry_CSR2015_2702, Gao_JPCC2015_13124}. As described above, the metal $5d$ orbitals in the hafnium compounds are initially unoccupied. Consequently, there is no reason to believe that the initial filling will result in a change of the atomic coordination. Instead, the change in the Bragg peak positions appears to be caused by the ordering of the potassium atoms which rearrange themselves depending on their concentration. Moreover, a structural phase change has not been reported or predicted in the literature. Calculations for lithium-intercalated ZrS$_2$, which closely resembles HfS$_2$, indicate no significant structural distortions \cite{Zhao_JPCC2019_2139}.

The intercalation affected the planar lattice parameters only slightly by changing them by not more than $\SI{1}{\%}$ (not shown). In contrast, the $c$-constants increased by $2.03-\SI{2.17}{\angstrom}$ ($33-\SI{36}{\%}$) which is approximately the thickness of one potassium layer. The details are listed in Table \ref{tab:cLatticeExp}. These percentage changes are similar to those observed in potassium-intercalated MoS$_2$ \cite{Habenicht_Sfp_, Somoano_TheJournalofChemicalPhysics_1973_58_2_697, Ren_NR2017_1313,Ruedorff_Chimia_1965_19__489, Habenicht_Sfp_} and TaSe$_2$ \cite{Koenig_EL2012_27002}. The data show that the expansion is slightly reversed at high doping levels, a fact also reported for tantalum diselenide \cite{Koenig_EL2012_27002}. The inter-planar widening is largely due to the size of the potassium atoms which move into the Van-der-Waals gaps spreading the comparatively rigid molecular crystal planes apart from each other. That process has a large effect on the layer spacing even at low alkali metal concentration $x$ but levels out quickly. However, the interlayer bonding grows with increasing doping levels because of the larger number of electrons in the conduction band. This counteracts the expansion at larger values of $x$. A systematic comparison between experimental and DFT lattice constants is presented in Fig. \ref{fig:CalcParameters}. The out-of-plane lattice parameters are largest for $x\approx0.25$ before they begin to contract again. The calculated expansion percentages agree very well with the experimentally observed values.

It should be mentioned that x-ray measurements performed by Whittingham and Gamble \cite{Whittingham_Mrb1975_363} found a unit cell spanning 3 molecular layers for lithium-intercalated HfS$_2$ which is in surprising contrast to our results. We assume that those studies were done on samples with low intercalation levels still showing some degree of disorder.
\subsection{Semiconductor-to-Metal Transition}\label{sec:STM_Transition}
\begin{figure*}
	\includegraphics []{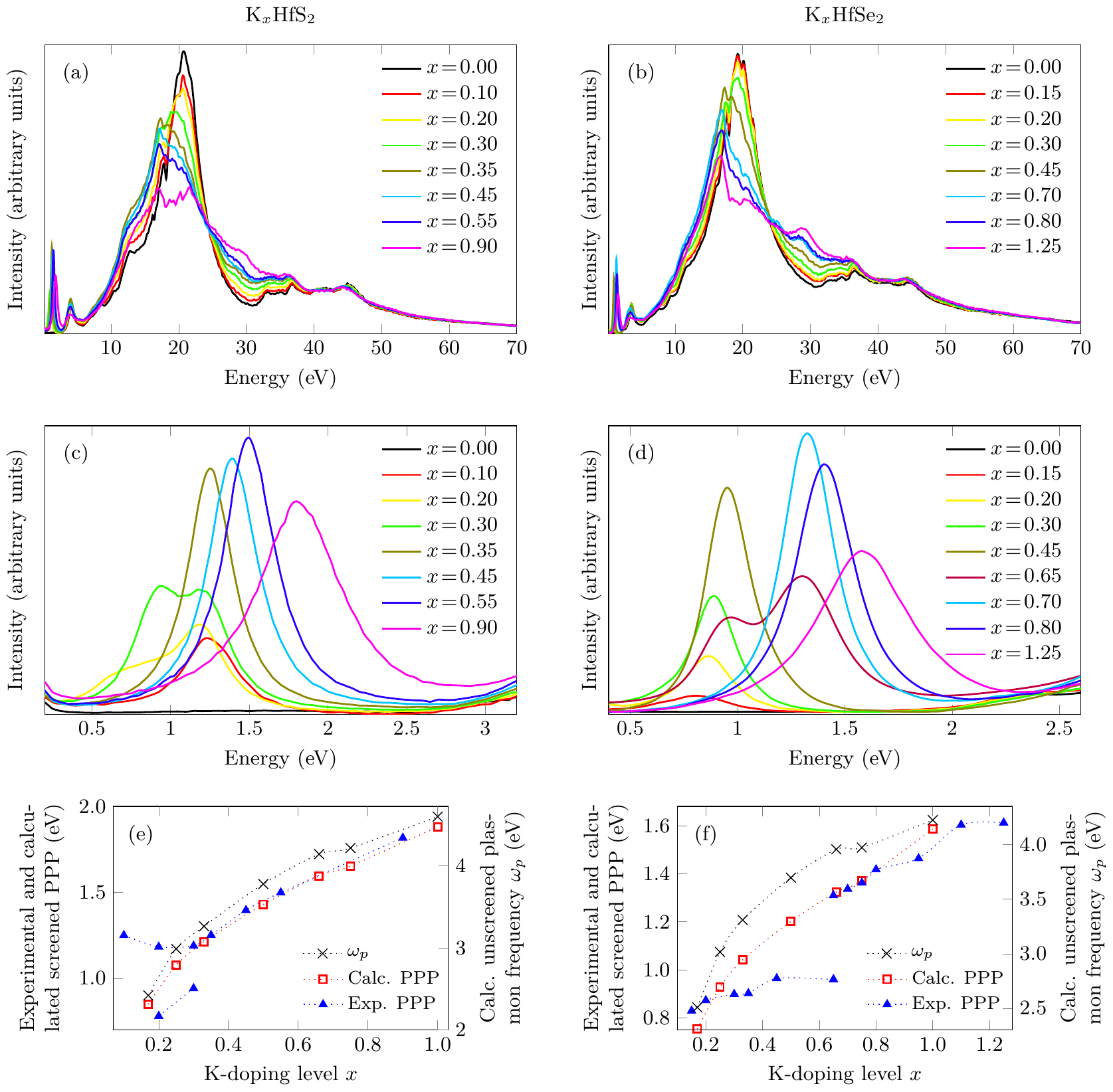}
	\caption{(Color online) (a) and (b): EELS spectra for pristine and potassium-intercalated HfS$_2$ and HfSe$_2$ for $|\mathbf{q}|=\SI{0.1}{\per\angstrom}$ parallel to the $\Gamma\,M$ direction at $\SI{20}{\K}$. The spectra were normalized at $\SI{70}{\eV}$. (c) and (d): Magnification of the spectra for the low-energy region. (e) and (f): Energy position of the plasmon peak (PPP) as a function of the potassium concentration extracted from the experimental and calculated energy-loss spectra as well as the calculated unscreened plasmon frequency $\omega_p$. The dotted lines serve as a guide for the eye.}
	\label{fig:DopingSteps}
\end{figure*}
Fig. \ref{fig:DopingSteps} (a) and (b) show the energy-loss spectra for hafnium disulfide and diselenide for a number of intercalation levels acquired at a momentum transfer value of $|\mathbf{q}|=\SI{0.1}{\per\angstrom}$ parallel to the $\Gamma M$-direction. Because of the isotropy of the spectra in all directions at such small |\textbf{q}|, it is not necessary to show the data for the $\Gamma K$-direction as well. The spectra are dominated by the volume plasmons, the collective oscillations of all valence electrons. They are centered around $\SI{20}{\eV}$ for the undoped compounds [black plots in Fig. \ref{fig:DopingSteps} (a) and (b)]. The features between $\SI{4}{\eV}$ and $\SI{12}{\eV}$ represent interband transitions \cite{Bell_AdvancesinPhysics_1976_25_1_53}. Hafnium $4f$ core level excitations account for the peaks at $\SI{17.6}{\eV}$ and $\SI{22.0}{\eV}$. There are also stimulations of the Hf $5p$ states between $\SI{33}{\eV}$ and $\SI{45}{\eV}$ in both materials \cite{Bell_AdvancesinPhysics_1976_25_1_53} that are superimposed on the effects of multiple scattering. For increasing doping levels, the volume plasmon peaks become more jagged and a feature forms at $\sim \SI{18}{\eV}$ arising from K $3p$ core levels.

In the low energy region, the spectra for the undoped crystals [black plots in Fig. \ref{fig:DopingSteps} (c) and (d)] display band gaps followed by excitonic transitions \cite{Habenicht_PRB2018_155204}. As potassium is added, new features begin to form initially around $\SI{1}{\eV}$. They shift to higher energies and rise in intensity as the K-concentration is increased before their intensity declines again while the peaks become broader. The fact that those new excitations develop in the energy region of the former band gaps and as a result of the intercalation with an electron donor suggests that they represent charge carrier plasmons and that semiconductor-to-metal transitions have occurred. The same phenomenon has been observed in K-intercalated 
WSe$_2$ \cite{Ahmad_JoPCM2017_165502} which is also a native semiconductor. The transition can also be seen in the shift of the Fermi energy in the calculated density of states leading to partially filled conduction bands (see Fig. \ref{fig:DOS-All} in Appendix \ref{sec:AppendixA}). 

Closer inspection of the DOS depicted in Fig. \ref{fig:DOS-All} reveals that, for all cases with $0 < x < 0.6$, the conduction band bottom is characterized by a jump-like onset followed by an almost constant DOS up to the Fermi level and beyond. This means, the related systems host a quasi-twodimensional (2D) electron gas at the Fermi level. Given the observed thermodynamic stability for $x > 0.3$, the title systems could form a platform for investigations on a 2D electron gas with densities of $2 \cdot 10^{14} - 6 \cdot 10^{14}$ electrons per cm$^2$. We note that, a quasi-2D electronic structure may seem natural for the given anisotropic structure. However, at higher doping levels, van-Hove singularities other than 2D-like signal a 3D electronic structure close to the Fermi level (Fig. \ref{fig:DOS-All} o, p).

Let us turn back our attention to the EELS data. It is unusual that such spectra assume the shape of a double peak as can be seen for HfS$_2$ at $x=0.20$ and 0.30. The two maxima are at $\SI{0.94}{\eV}$ and $\SI{1.19}{\eV}$ [Fig. \ref{fig:DopingSteps} (c)]. For HfSe$_2$ they are located at $\SI{0.94}{\eV}$ and $\SI{1.31}{\eV}$ for $x=0.65$ [Fig. \ref{fig:DopingSteps} (d)]. Moreover, no plasmon peak forms in the energy region between the two peaks for any of the investigated doping levels. All other peak maxima are located either below or above those energies. This behavior can also be seen in Fig. \ref{fig:DopingSteps} (e) and (f) where the plasmon peak positions (PPP) are plotted against the doping concentrations. In HfS$_2$, the energetically higher peak forms first before the double feature appears at higher $x$. In contrast, the energetically lower lying peak develops before the occurrence of the second one in hafnium diselenide. This raises the question why plasmon formation is not observed in the energy range between the two peaks. Possible reasons could be a phase change at particular doping levels or certain arrangements of the potassium ions in the host lattices. Besides the peak-splitting, the plasmon energies appear to remain relatively constant until a certain K-concentration is exceeded after which the peak energy positions increase. 

\begin{figure}[!b]
	\includegraphics [width=0.47\textwidth]{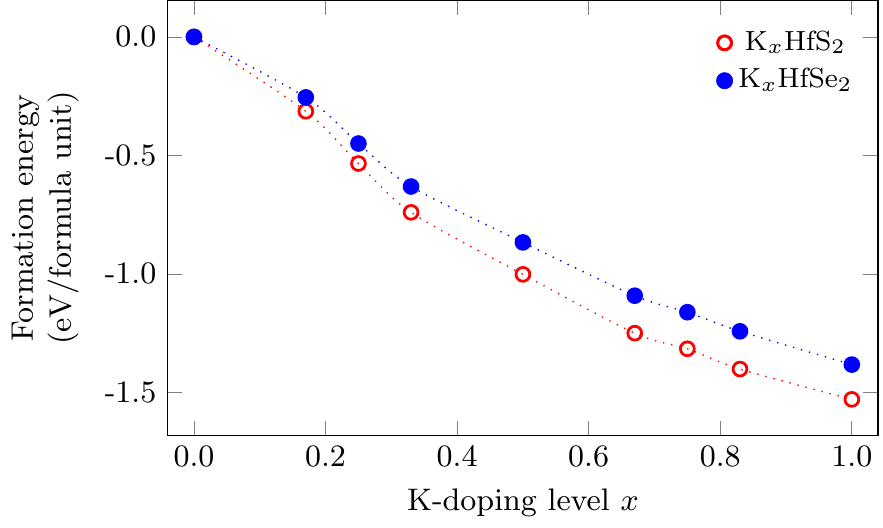}
	\caption{(Color online) Formation energies for K$_x$HfS$_2$ and K$_x$HfSe$_2$ for various potassium doping levels.}
	\label{fig:FormEnergies}
\end{figure}

To gain a better understanding of the reasons for those two observations, we calculated the formation energies per formula unit (f.u.) $E_{\rm form}$ of the compounds for selected alkali metal concentrations.
\begin{equation}
E_{\rm form} = E_{\rm K_xHfCh_2} - x E_{\rm bcc-K} - E_{\rm HfCh_2} \; ,
\label{equ:Formation}
\end{equation}
with $E_{\rm K_xHfCh_2}$ denoting the energy per f.u. of the doped chalcogenide (Ch = S, Se); $E_{\rm bcc-K}$ and $E_{\rm HfCh_2}$ denoting the energies per f.u. of the reference systems bcc Potassium and HfCh$_2$, respectively. The results are summarized in Fig. \ref{fig:FormEnergies}. The concave slope of the plots for doping levels up to $x \approx 0.3$ indicates that a homogeneous doped phase is unstable below that potassium concentration. In contrast, the slope is convex for higher alkali metal concentrations implying that all structures considered in the simulations with $x \gtrsim 0.3$ are low-temperature stable against decomposition into 
structures with different doping levels. We note that the DFT calculations were carried out for bulk systems. Experimental data were obtained for films and could be slightly influenced by surface/interface effects, or by kinetics.
\begin{figure*}
	\includegraphics [width=\textwidth]{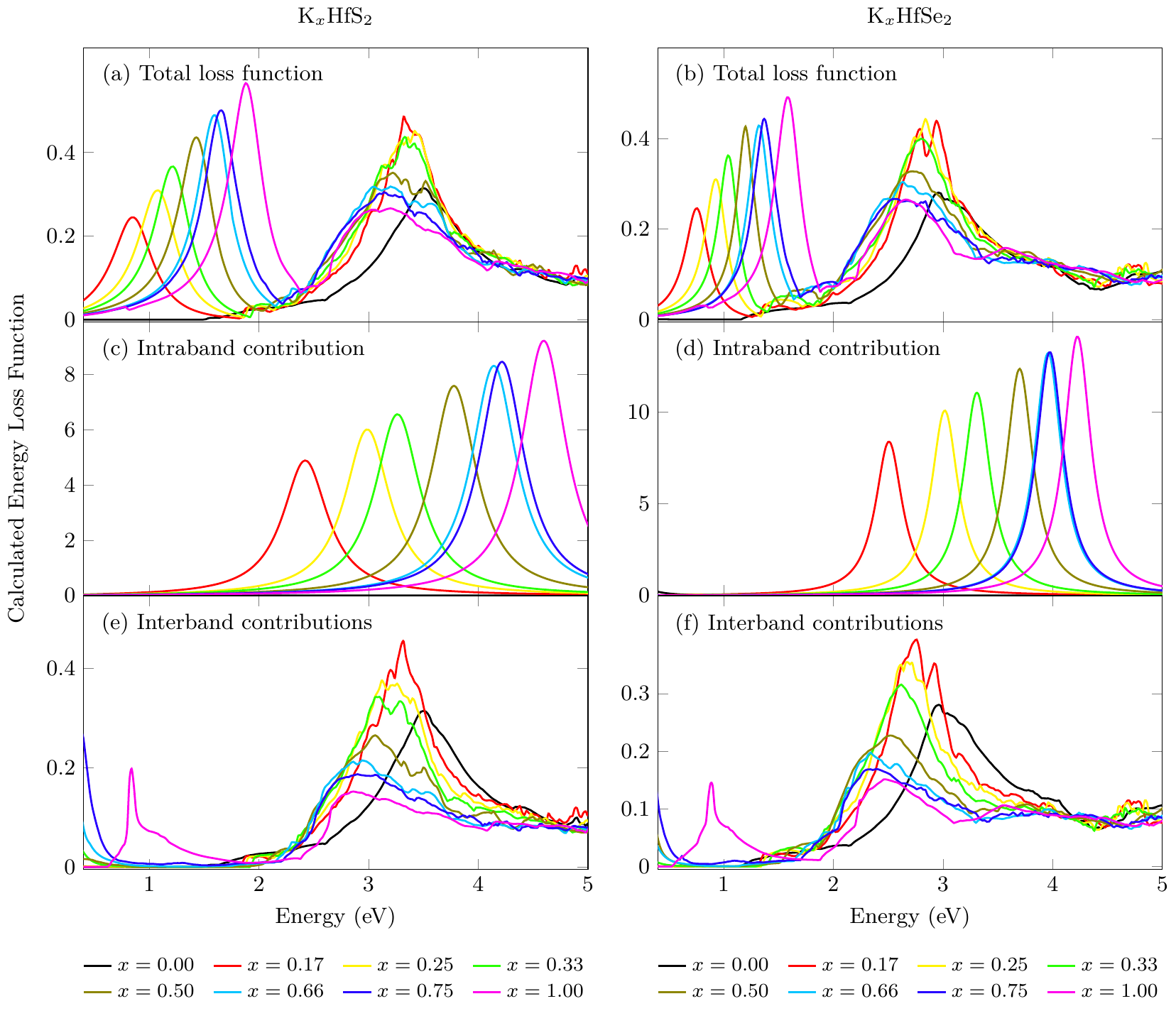}
	\caption{(Color online) (a) and (b): Calculated in-plane loss functions for K$_x$HfS$_2$ and K$_x$HfSe$_2$ for various doping levels. (c) and (d): Calculated intraband contributions to the in-plane loss functions. The frequency widths (Equ. \ref{equ:epsilon}) used in the calculations were $\gamma=\SI{0.5}{\eV}$ and $\SI{0.3}{\eV}$ for HfS$_2$ and HfSe$_2$, respectively. (e) and (f): Calculated interband contributions to the in-plane loss functions.}
	\label{fig:CalcDepSeries}
\end{figure*}
The suggested thermodynamic instability of low potassium concentrations could explain the fact that the plasmon position is relatively unchanged during the initial doping steps. If the amount of alkali metal is not sufficient to saturate the whole crystal uniformly at a thermodynamically stable concentration, domains will form to accommodate the potassium at the smallest stable concentration. As more potassium is added, the volume of the already intercalated regions increases at the expense of the pristine domains whose volume shrinks. The potassium concentration and, consequently, the density of the supplied conduction electrons remains constant in the intercalated domains during this process. This conduction electron density $n$ determines the unscreened plasmon frequency $\omega_p$ according to Equ.~\ref{equ:PlasmonFrequDrude}. The screened plasmon frequency changes almost proportionally to $\omega_p$. Their relation can be seen by comparing the positions of screened and unscreened plasmon frequencies obtained from DFT calculations in Fig.~\ref{fig:DopingSteps} (e) and (f). Consequently, the experimental plasmon peak position, which is close to the screened plasmon frequency, changes almost proportionally to the square root of the charge carrier density. Therefore, the relative stability of the plasmon peak position in the EELS spectra is an indication of a constant potassium concentration. In K$_x$HfS2$_2$, this is the case up to $x\approx 0.35$ and in K$_x$HfSe$_2$ up to $x\approx 0.65$. Before those points, the calculated potassium concentrations $x$ represent the average concentrations across the whole samples and not the concentration in the intercalated domains. The expansion of the intercalated domains leads to an enhancement of the plasmon intensities in Fig.~ \ref{fig:DopingSteps} (c) and (d). Once the whole film has reached the minimum stable concentration, the doping level and the plasmon energy position increase smoothly as more potassium is provided. It is interesting that in contrast to those observations, WSe$_2$ \cite{Ahmad_JoPCM2017_165502}, K$_x$CuPc \cite{Flatz_TJocp2007_214702}, and K$_2$MnPc \cite{Mahns_TJocp2011_194504} permit only one particular potassium stoichiometry causing the plasmon peak position to be almost unchanged during the intercalation steps. On the other hand, the metallic TMDCs TaSe$_2$, TaS$_2$, NbSe$_2$ and NbS$_2$ appear to accept any alkali metal concentration \cite{Koenig_EL2012_27002, Mueller_PRB2016_35110}.

We used the FPLO code to calculate the energy loss spectra for different doping levels to determine if they would reproduce the experimental results.
The calculated plots, which are presented in Fig. \ref{fig:CalcDepSeries} (a) and (b), show a single peak moving to higher energies with increasing doping level. A peak splitting of the kind seen in the measured data cannot be identified. The emergence of the spectral double features at certain doping levels most likely arises from the temporary formation of two domains with differing doping concentrations that depart significantly from the thermodynamic equilibrium. Those domains are different from the ones described in the preceding paragraph which exist in equilibrium conditions. This additional kinematic effect may be caused by the experimental process where just one side of the crystal is exposed to the potassium stream during the intercalation leading to initially inhomogeneous alkali metal distributions. Such effects cannot entirely be prevented even though the samples were annealed after each intercalation step to minimize such issues. This reasoning is supported by the fact that the energetically higher one of the two peaks in K$_x$HfSe$_2$ at $x=0.65$ disappeared after longer electron beam exposure. The energy supplied by the beam may have induced a further migration of the potassium atoms and a more uniform distribution resulting in an equalization of the two regions. Another observation corroborating this assumption is that the double features exist only for doping levels for which the crystals display significant disorder in the $c$-direction (see Sec. \ref{sec:DiffractionPatterns}).

We computed the intraband and interband contributions to the loss function using FPLO. The intraband parts reflect the unscreened plasmons as presented in Fig. \ref{fig:CalcDepSeries} (c) and (d). Just like the screened plasmon peaks, the unscreened plasmon features shift to higher energies with increasing $x$ and show no unusual behavior. For the undoped materials, the interband contributions in Fig. \ref{fig:CalcDepSeries} (e) and (f) exhibit the expected band gaps followed by the exciton signatures. At $x=0.75$, weak interband transitions begin to emerge near $1.1 - \SI{1.3}{\eV}$. Stronger excitations occur close to $\SI{0.8}{\eV}$ for $x=1.00$. Nevertheless, the interband excitations are relatively weak compared to the intraband excitations. This verifies that the peaks in the loss function below $\sim \SI{2}{\eV}$ are largely plasmonic in nature and that the same is true for the corresponding features in the experimental loss spectra.

\subsection{Unscreened Charge Carrier Plasmon Frequency}\label{sec:PlasmonFrequency}
\begin{figure}
	\includegraphics [width=0.48\textwidth]{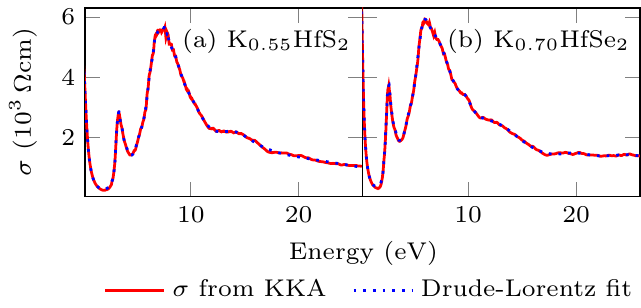}
	\caption{(Color online) Optical conductivity based on the Kramers–Kronig analysis of the electron-energy loss spectrum and its fitting with the Drude-Lorentz model for (a) K$_{0.55}$HfS$_2$ and (b) K$_{0.70}$HfSe$_2$.}
	\label{fig:KKA-Fits}
\end{figure}
Because of their significance, we calculated the unscreened charge carrier plasmon frequencies from the measured EELS data. To extract the experimental values, it was necessary to separate the plasmons from the single particle excitations in the EELS spectra. For that purpose, the data for K\textsubscript{0.55}HfS$_2$ and for K\textsubscript{0.70}HfSe$_2$ measured in the energy range up to $\SI{100}{eV}$ parallel to the $\Gamma M$ direction were corrected for experimental artifacts by eliminating the elastic line, centered at $\SI{0}{\eV}$, and the effects of multiple scatting according to the approach outlined in Refs. \citenum{Fink_AEEP_1989_75__121} and \citenum{Schuster_PRB2009_45134}. A Kramers-Kronig analysis was performed on the outcomes based on the assumption that the samples were metallic. The resulting optical conductivity function $\sigma(\omega)$ [$=\epsilon_0\omega\epsilon_{im}(\omega)$] is plotted in Fig. \ref{fig:KKA-Fits}. It was fitted with one Drude and 14 Lorentz oscillators in the energy region up to $\SI{26}{eV}$ to obtain the parameters in Equ. \ref{equ:epsilon}. The fitted values of $\omega_{p}$ stopped fluctuating for a larger number of oscillators. The complete fit parameter sets are provided in Table \ref{tab:FitParameters} in Appendix \ref{sec:AppendixA}. The plots produced from them are displayed in Fig. \ref{fig:KKA-Fits}. They show that the Drude-Lorentz model provides a good description of the optical conductivities. They also indicate that $\sigma(\omega)$ below $\sim\!\SI{1}{eV}$ is dominated by the charge carrier plasmon (Drude oscillator) while excitations of bound single particles (Lorentz oscillators) account mainly for the behavior at higher energies. The process lead to unscreened plasmon frequencies of $\omega_p=\SI{3.55}{eV}$ for K$_{0.55}$HfS$_2$ and $\SI{3.89}{eV}$ for K$_{0.70}$HfSe$_2$, respectively. Those values are somewhat lower than the theoretical numbers for comparable doping levels of $\SI{3.86}{eV}$ for K$_{0.50}$HfS$_2$ and $\SI{4.14}{eV}$ for K$_{0.66}$HfSe$_2$. However, the values are in a similar energy range and reasonably close given the approximations we have made.

\subsection{Plasmon Dispersion}\label{sec:PlasmonDispersion}
\begin{figure}
	\includegraphics [width=0.48\textwidth]{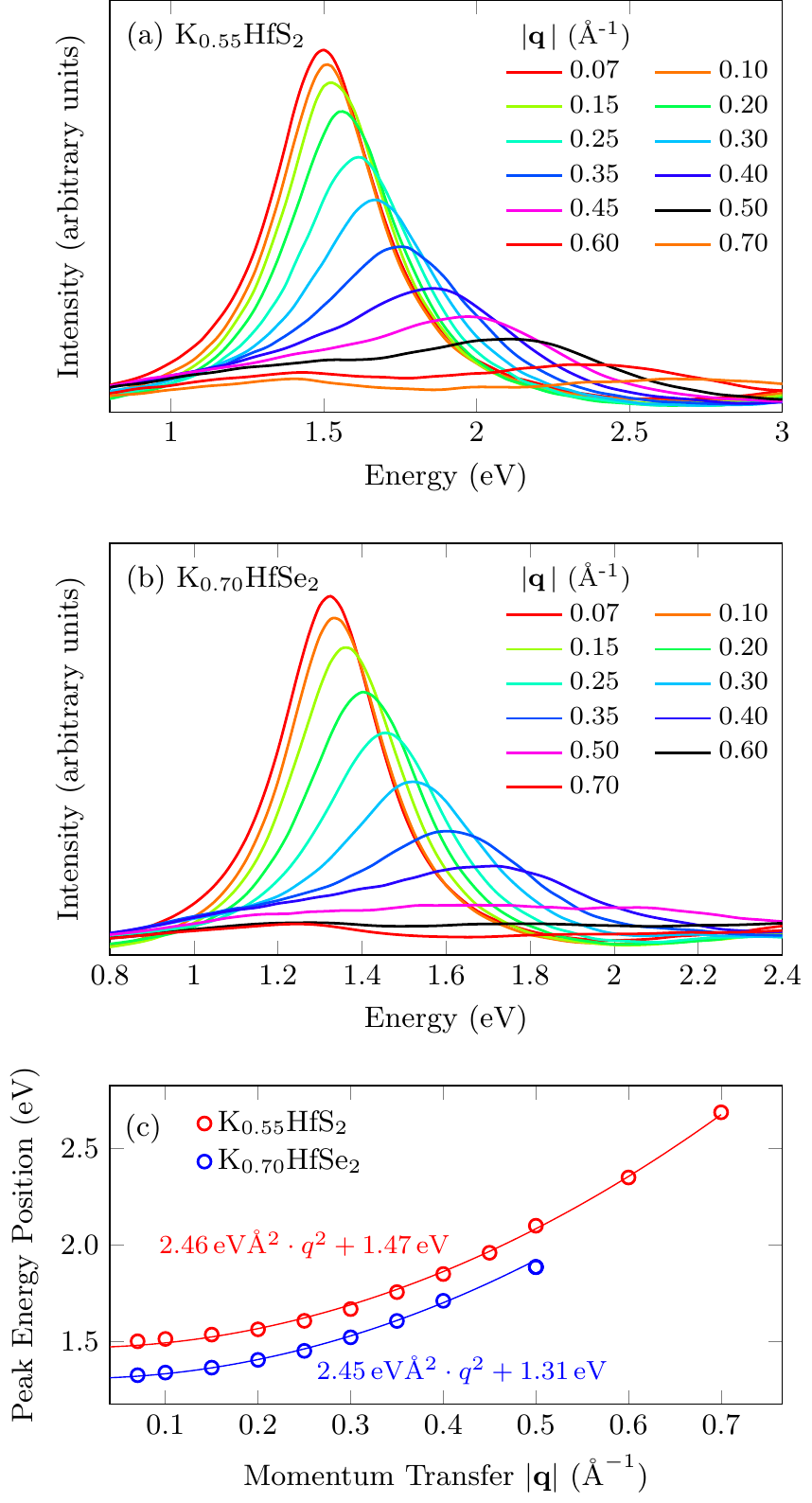}
	\caption{(Color online) EELS spectra of (a) K$_{0.55}$HfS$_2$ and (b) K$_{0.70}$HfSe$_2$ measured for the indicated momentum transfer values (|\textbf{q}|) in the $\Gamma K$-direction at $T=\SI{20}{\K}$. (c) Energy-momentum dispersions for the peaks shown in subfigures (a) and (b). For K$_{0.70}$HfSe$_2$, the plasmon energy peak positions cannot be identified for $|\mathbf{q}|>0.5 {\rm \AA}^{-1}$. The solid lines outline the quadratic fit of the dispersion.}
	\label{fig:Dispersion}
\end{figure}
Another interesting aspect of doping-induced plasmons is their dispersion. As stated above, the energy position of a plasmon peak in a spectrum corresponds to the unscreened plasmon frequency $\omega_p$ damped mainly by interband excitations. 
Besides an energy renormalization, the fundamental behavior of the screened plasmon peak position and the unscreened plasmon peak position is nearly the same \cite{Mueller_PRB2017_75150}. This allows us to use the dispersion of the measured peak position as an approximation of the momentum dependence of the unscreened plasmon frequency. In an ideal metal, this frequency, and therefore the plasmon peak position, changes almost quadratically as function of momentum \cite{Nolting_2018_695, Nozieres_PR1959_1254}:
\begin{equation}
	\omega_p(|\mathbf{q}|)\approx\omega_p(0)+\frac{3}{10}\frac{\hbar^2k_f^2}{m^{*2}_e\omega_p(0)}|\mathbf{q}|^2+O(|\mathbf{q}|^4)
	\label{equ:PlasmonDisp}
\end{equation}
where $\hbar$ represents the Planck constant and $k_f$ the Fermi wave vector. However, experiments also found results that deviate from this ideal behavior. For example, the energy-momentum relations in TaS$_2$, TaSe$_2$ and NbSe$_2$ are negative \cite{Schuster_PRB2009_45134, Wezel_PRL2011_176404, Mueller_PRB2016_35110}. They become positive and linear upon alkali metal intercalation \cite{Mueller_PRB2016_35110, Koenig_PRB2013_195119}. Bi$_2$Sr$_2$CaCu$_2$O$_8$, on the other hand, has quadratic dispersion \cite{Nuecker_PRB1991_7155, Grigoryan_PRB1999_1340}.

For each of the two materials under investigation, the doping levels with the most intense plasmon peaks ($x=0.55$ for HfS$_2$ and $x=0.70$ for HfSe$_2$) were selected for an analysis of their momentum dependence. The spectra, measured for a range of momentum transfer values, are shown in Fig. \ref{fig:Dispersion} (a) and (b) and look very similar for both compounds. The peaks continuously shift to higher energies (up to $\sim \SI{2.69}{\eV}$ and $\SI{1.98}{\eV}$ for HfS$_2$ and HfSe$_2$, respectively) and broaden as |\textbf{q}| increases. The plots of the peak energy positions vs. momentum transfer in Fig. \ref{fig:Dispersion} (c) reveal a quadratic dispersion which coincides with the expectations for plasmons in ideal metals according to Equ. \ref{equ:PlasmonDisp}.

Besides that strongly dispersing feature, a shoulder begins to emerge in the spectra for HfS$_2$ near $\SI{1.45}{\eV}$ for $|\mathbf{q}|=0.40$ [Fig. \ref{fig:Dispersion} (a)]. It develops into a separate peak at higher momentum transfer values and has a slight negative dispersion. The same phenomenon occurs in HfSe$_2$ around $\SI{1.25}{\eV}$ [Fig. \ref{fig:Dispersion} (b)]. Those excitations represent interband transitions as can be seen from the calculated interband contributions to the loss functions in Fig. \ref{fig:CalcDepSeries} (e) and (f). The latter exhibit such transitions in the energy region between $\SI{0.8}{\eV}$ and $\SI{1.5}{\eV}$ for higher potassium levels. Their intensities are weak so they cannot be distinguished from the stronger plasmons at low |\textbf{q}|. As the momentum transfer is raised, the plasmon peaks themselves peter out and shift away revealing the less dispersive single particle transitions.

\subsection{ARPES Spectra}\label{sec:ARPES_Spectra}
\begin{figure*}
	\includegraphics [width=0.48\textwidth]{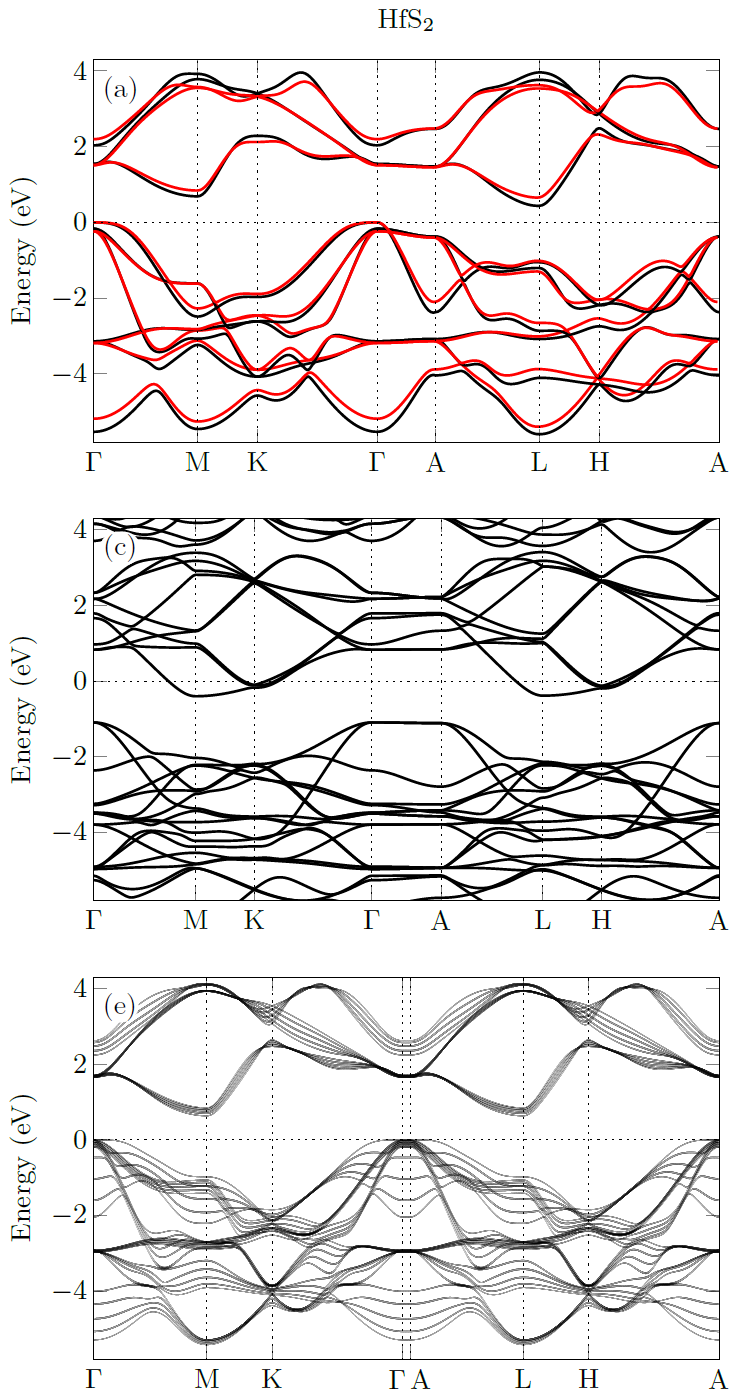}%
	\includegraphics [width=0.48\textwidth]{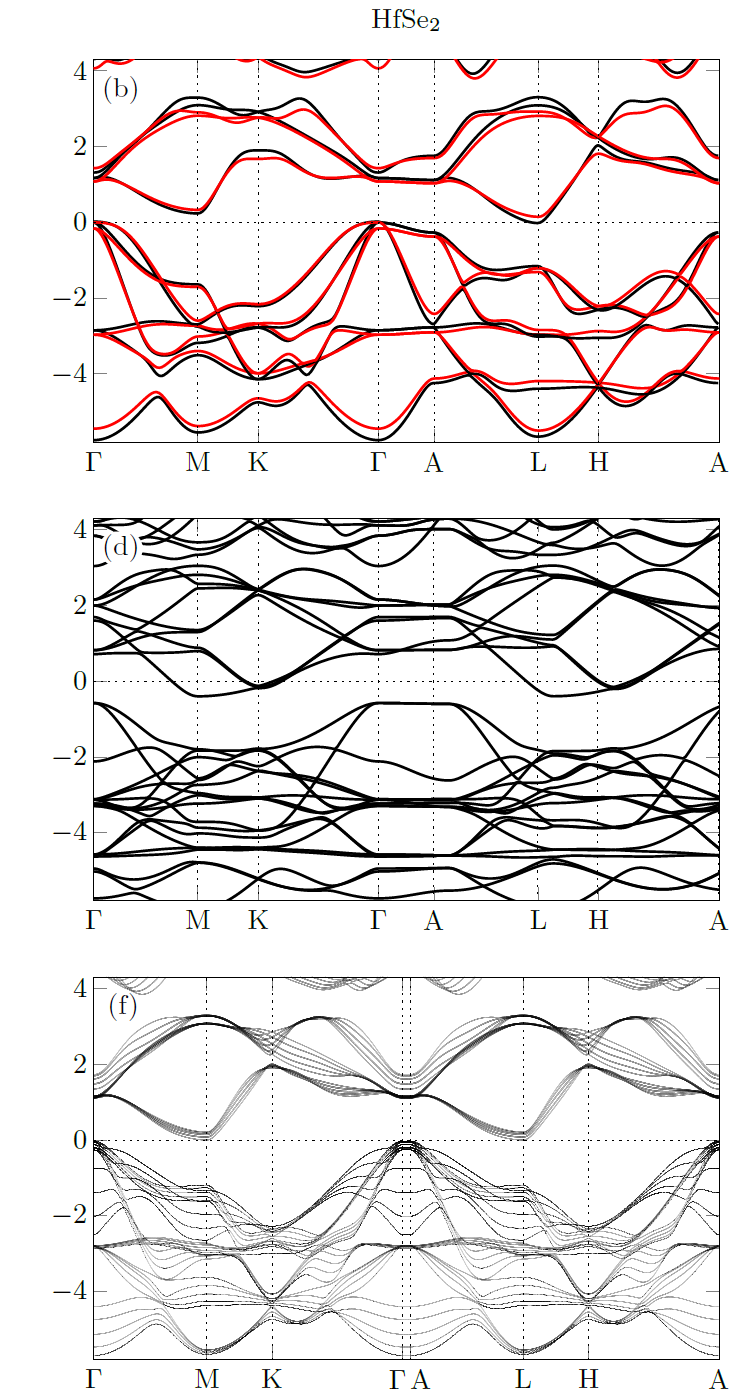}%
	\caption{(Color online) (a) and (b): LDA band structures for pristine bulk HfS$_2$ and HfSe$_2$, respectively, based on LDA-optimized (black lines) and experimental (red lines) lattice parameters. (c) and (d): LDA band structures for bulk K$_{0.33}$HfS$_2$ and K$_{0.33}$HfSe$_2$, respectively, based on LDA-optimized lattice parameters. (e) and (f): LDA band structures for a slab of 5 molecular layers (based on LDA-optimized unit cell parameters) of pristine HfS$_2$ and HfSe$_2$, respectively. The thickness of the band plots indicates the contributions of the photoelectrons to the spectral intensity based on their escape depth.}
	\label{fig:CalBS}
\end{figure*}
\begin{figure}
	\includegraphics [width=0.48\textwidth]{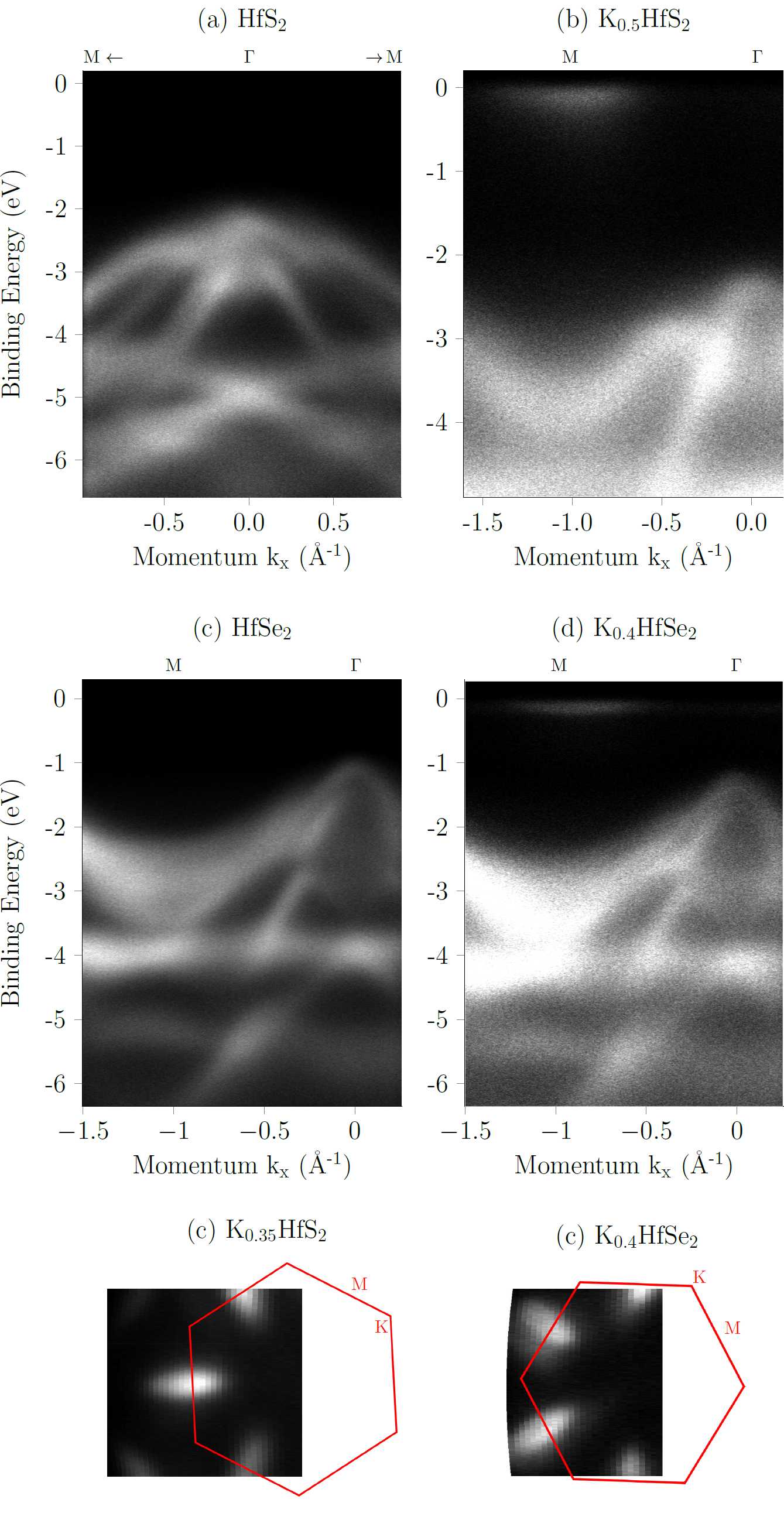}
	\caption{(Color online) (a) and (c): ARPES energy distribution curves for pristine HfS$_2$ and HfSe$_2$, respectively. (b) and (d): ARPES energy distribution curves for potassium-intercalated K$_{0.5}$HfS$_2$ and K$_{0.4}$HfSe$_2$, respectively. (e) and (f): In-plane ARPES spectra for K$_{0.35}$HfS$_2$ and K$_ {0.4}$HfSe$_2$, respectively, at the Fermi energy.}
	\label{fig:EDCs}
\end{figure}

To confirm the occurrence of the semiconductor-to-metal transition and the validity of our computational results, we performed ARPES measurements on pristine and potassium-doped hafnium disulfide and diselenide. Similar experiments have been done before on pure \cite{Aretouli_APL2015_143105} and sodium-doped \cite{Mleczko_Sa2017_1700481, Eknapakul_PRB2018_201104} HfSe$_2$.

We calculated the bulk band structures for the undoped materials. Fig. \ref{fig:CalBS} (a) and (b) display the results obtained from LDA-optimized as well as experimental lattice parameters (see Sec. \ref{sec:DiffractionPatterns} for the discussion of the experimentally determined lattice parameters). The two data sets show some moderate differences in the energy positions of parts of certain bands but no fundamental differences regarding the relative locations of the main features. The band structure for HfSe$_2$ with optimized lattice parameters appears to be metallic even though the material is a semiconductor. 
The valence band maxima (VBM) are located at the $\Gamma$ point which agrees with theoretical results found by others \cite{Murray_JoPCSSP1972_746, Mattheiss_PhysicalReviewB_1973_8_8_3719, Fong_PRB1976_5442,Traving_PRB2001_35107, Reshak_PBCM2005_25,Jiang_TJoCP2011_204705}. According to our calculations, the conduction band minima (CBM) are at the $L$ point. Because the local conduction band minima at the $L$ and $M$ points are energetically very close, it has been debated which one of them represents the absolute CBM with some calculations pointing to $L$ \cite{Jiang_TJoCP2011_204705,Mattheiss_PhysicalReviewB_1973_8_8_3719}, some to M \cite{Fong_PRB1976_5442, Reshak_PBCM2005_25}, and some undetermined results \cite{Traving_PRB2001_35107}. Photoemission \cite{Traving_PRB2001_35107} and optical transmission \cite{Terashima_SSC1987_315} experiments suggest that the indirect gap is between $\Gamma$ and $L$. 
The direct gap of HfS$_2$ is in the range of $2.1-\SI{2.87}{\eV}$ according to conductivity and reflectivity measurements \cite{Bayliss_JoPCSSP1982_1283, Conroy_IC1968_459} while absorption and differential transmission experiments report indirect band gaps of $1.75 -\SI{2.26}{\eV}$ \cite{Camassel_INCB11977_185, Gaiser_PRB2004_75205, Greenaway_JPCS1965_1445, Yacobi_JoPCSSP1979_2189}.
For HfSe$_2$, reflectivity experiments observed a direct gap of $\SI{2.02}{\eV}$ \cite{Bayliss_JoPCSSP1982_1283}. Absorption and scanning tunneling spectroscopy indicate an indirect gap of $1.1 - \SI{1.13}{\eV}$ \cite{Yue_AN2015_474, Greenaway_JPCS1965_1445, Gaiser_PRB2004_75205}.
Our calculations result in significantly smaller gaps. This is not unexpected given that LDA tends to underestimate those values \cite{Aryasetiawan_RoPiP1998_237}.

The theoretical bulk band structures [Fig. \ref{fig:CalBS} (a) and (b)] deviate in some details from the ARPES spectra of the pristine samples in Fig. \ref{fig:EDCs} (a) and (c). Note that the energy zero is placed at VBM in the calculated band structures. For example, the topmost band from the theoretical spectra appears to be shifted to a lower energy in the photoemission spectra leading to an additional shoulder. Moreover, there seem to be a number of diffuse, indistinguishable bands at $\Gamma$ between $\SI{-3.5}{\eV}$ and $\SI{-2.5}{\eV}$ for the sulfide compound and between $\SI{-3.5}{\eV}$ and $\SI{-1.5}{\eV}$ for the selenide one. 
One of the main factors contributing to the differences is that ARPES experiments are surface sensitive while the plots in Fig. \ref{fig:CalBS} (a) and (b) were derived from calculations for bulk materials. To simulate the fact that the photoelectrons can escape only from positions very close to the sample surface, the band structures for a periodic slab of 5 molecular layers surrounded by vacuum were determined and presented in Fig. \ref{fig:CalBS} (e) and (f). The bands are weighed by the distance of the photoelectron source from the sample surface according to Equ. \ref{equ:IntDecay} with an escape depth of $\SI{4}{\angstrom}$. The resulting spectra resemble the experimental outcomes much better, in particular the valence band dispersion around $\Gamma$ and the multitude of bands below the VBM seen in the measured data. The outcome compares well to calculations performed by Aretouli \textit{et al.} for free-standing 6-layer HfSe$_2$ \cite{Aretouli_APL2015_143105}.
Nevertheless, the use of the bulk calculations is appropriate for the interpretation of the transmission EELS spectra because this method is bulk-sensitive.

The ARPES spectra for the K-doped samples are depicted in Fig. \ref{fig:EDCs} (b) and (d). For $x\approx0.5$, the influx of electrons from potassium atoms leads to a shift of the valence band maximum at $\Gamma$ from $\sim\SI{2.1}{\eV}$ to $\sim\SI{2.3}{\eV}$ below the Fermi energy in HfS$_2$ indicating a rise of $E_F$ by $\sim\SI{0.2}{\eV}$. The shift creates an electron pocket at the Fermi energy representing the minimum of the now partially filled conduction band. 
The indirect gap between the VBM at $\Gamma$ and the CBM at $M$ amounts to $\SI{2.2}{\eV}$. The position of the VBM is consistent with the calculated VBM at the $M/L$ points [see Fig. \ref{fig:CalBS} (c)] obtained for an alkali metal concentration of $x=0.33$, which is reasonably close to the actual doping level. The observation of this Fermi pocket is a clear sign of the semiconductor-to-metal transition.

Similarly, the energy maximum at $\Gamma$ decreases by $\SI{0.3}{\eV}$ to $\SI{-1.3}{\eV}$ in HfSe$_2$ for $x\approx0.4$. The energy difference between the electron pocket at $M$ and the VBM is $\SI{1.2}{\eV}$. This is slightly lower than Mleczko \textit{et al.} observed for an ARPES investigation of sodium-intercalated HfSe$_2$ \cite{Mleczko_Sa2017_1700481}. As stated above, the calculated LDA band gaps in Fig. \ref{fig:CalBS} (c) and (d) are smaller than the experimental values.
The hexagonal arrangement and shape of electron pockets is visible in the momentum distribution curves shown in Fig. \ref{fig:EDCs} (a) or (b).

\section{SUMMARIZING DISCUSSION}
We used transmission electron energy loss spectroscopy and angle-resolved photoemission spectroscopy supported by DFT calculations to investigate the effect of potassium intercalation on bulk single crystals of HfS$_2$ and HfSe$_2$. Electron diffraction patterns showed a significant degree of disorder in the crystal structures for low doping concentrations. The structures become well-ordered again at alkali metal levels of $x_{\mathrm{HfS_2}}=0.55$ and $x_{\mathrm{HfSe_2}}=0.70$. At those points the materials show in-plane parameter changes of less than $\SI{1}{\%}$ and out-of-plane lattice expansions of $33-\SI{36}{\%}$. Calculations indicate that the latter expansions reach their maximum at $x\approx0.25$ before they begin slightly to retract. Moreover, superstructures appear in the planes that we attribute to an ordered arrangement of the potassium ions minimizing their electrostatic (Madelung) energy. 

The intercalation leads to the formation of a new feature in the energy-loss spectra below $\SI{2}{\eV}$ which can be identified as a charge carrier plasmon based on the calculated intraband contribution to the loss functions. It is a clear indication of a semiconductor-to-metal transition supported by computed DOSs and band structures. Its peak position remains relatively stable up to a certain potassium load. Close to this load, a double peak is observed and the doping-level dependence of the peak positions shows a clear gap. Related DFT calculations of the formation energies show that low potassium concentrations are thermodynamically unstable.
These two observations indicate the formation of domains at low potassium load, i.e., the pristine phase coexists with a phase of lowest stable doping level $x$. A possible reason for the instability of a low-$x$ phase could consist in the almost $x$-independent effort to separate adjacent HfS$_2$ or HfSe$_2$ layers to accommodate potassium atoms which is counterbalanced  at higher $x$ by a gain in the binding energy of potassium that is roughly proportional to $x$. Yet higher doping concentrations results in a convex formation energy, Fig. \ref{fig:FormEnergies}, due to growing electrostatic repulsion among the dopands.
As soon as a sufficient amount of potassium is intercalated to saturate the hole crystal with the minimum stable concentration, the potassium stoichiometry increases continuously for subsequent intercalation steps. This is reflected in a square-root like increase of the plasmon energy position. 

The plasmons exhibit a quadratic momentum dispersion which leads to the revelation of weak interband transitions in the same energy region at larger $q$-values. 

ARPES measurements on the intercalated compounds show electron pockets from the conduction band corroborating the transition to metallic behavior. 

Inspection of the calculated DOS and band structure revealed that the conduction band bottom hosts an almost ideal 2D electron gas for $x\approx0.5$, with an approximate density of $4\cdot 10^{14}$ electrons per cm$^2$.

\begin{acknowledgments}
We thank R. H\"ubel, S. Leger, M. Naumann, F. Thunig and U. Nitzsche for their technical assistance. R. Schuster and C. Habenicht are grateful for funding from the IFW excellence program.
\end{acknowledgments}

\clearpage
\section{APPENDIX} \label{sec:AppendixA}

\begin{figure}[h]
	\includegraphics [width=0.45\textwidth]{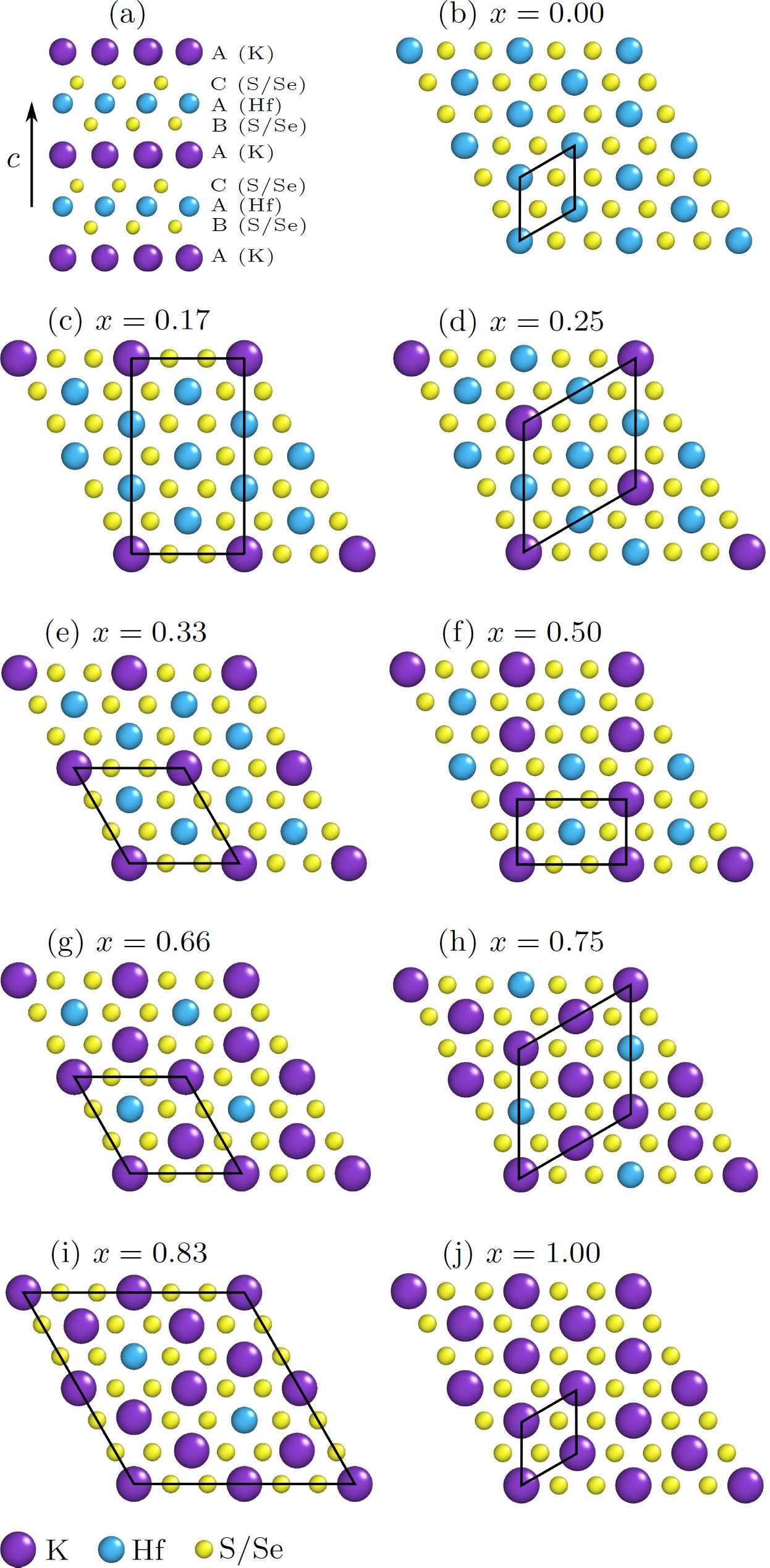}
	\caption{(Color online) (a) Out-of-plane crystal structure showing the arrangement of the crystal layers chosen for the calculations including the alignment of the intercalated potassium atoms in the Van-der-Waals gaps. The atomic plane sequence is defined to the right of the picture. (b) - (j): $ab$-planes of crystal structures demonstrating the chosen distribution of the potassium atoms for the various doping levels $x$ used in the calculations. The unit cells/supercells are outlined by black parallelograms.}
	\label{fig:Supercells}
\end{figure}
\clearpage
\setlength\LTcapwidth{\textwidth}
\begin{longtable*}{l>{\centering\arraybackslash}p{0.1cm}c>{\centering\arraybackslash}p{0.1cm}>{\centering\arraybackslash}p{2.6cm}>{\centering\arraybackslash}p{2.6cm}>{\centering\arraybackslash}p{2.6cm}}
\caption{Optimized atomic coordinates, experimental/optimized lattice parameters, space groups and $k$-meshes for K$_x$HfS$_2$. The $k$-meshes were chosen such that the $k$-point density is approximately the same in all structures.}
\label{tab:HfS2_CalcDetails}\\
\hline\hline
					&&&\multicolumn{4}{c}{Optimized atomic coordinates}\\ \cline{5-7}
Calculation parameters	&&Atom 	&&x	/a	&y/b	&z/c \\ 
\hline
\endhead
\hline\hline
\endfoot
\\
\underline{1. Bulk HfS$_2$ (experimental lattice parameters):}	\\
Space group: $P\bar3m1$ (164)											&&Hf	&&0		&0		&0	\\
$k$-mesh for structure optimization: $24\times24\times13$						&&S		&&1/3	&-1/3	&0.248	\\
$k$-mesh for band structure/DOS/Optics: $52\times52\times28$	\\
Lattice parameters: $a=b=\SI{3.63}{\angstrom},\,c=\SI{5.77}{\angstrom}$ \\
\\
\underline{2. Bulk HfS$_2$ (optimized lattice parameters):}					\\ 
Space group: $P\bar3m1$ (164)											&&Hf		&&0		&0		&0	\\
$k$-mesh for structure optimization: $24\times24\times13$						&&S		&&1/3	&-1/3	&0.256	\\
$k$-mesh for band structure/DOS/Optics: $52\times52\times28$		\\
Lattice parameters: $a=b=\SI{3.56}{\angstrom},\,c=\SI{5.71}{\angstrom}$ \\ 
\\
\underline{3. Bulk K$_{0.17}$HfS$_2$ (optimized lattice parameters):}\\
Space group:	 $P1\,2/m1$ (10)												&&Hf		&&0		&0		&0	\\
	$k$-mesh for structure optimization: $12\times8\times13$					&&Hf		&&0		&-1/2		&-1/2	\\
	$k$-mesh for band structure/DOS/Optics: $25\times18\times27$				&&Hf		&&0		&-0.336		&0	\\
Lattice parameters: $a=\SI{6.15}{\angstrom},\,b=\SI{10.65}{\angstrom},\,c=\SI{7.62}{\angstrom}$	&&Hf	&&0		&0.170		&-1/2	\\
Axis	angles: $\alpha=\beta=\gamma=\SI{90}{\degree}$						&&S		&&-0.193	&-0.167	&-0.167	\\
																		&&S		&&0.194		&-0.333	&-0.333	\\
																		&&S		&& -0.194	&-1/2	&-0.167	\\
																		&&S		&&-0.192	&0	&0.335	\\
																		&&K		&&1/2		&0		&0	\\
\\ 
\underline{4. Bulk K$_{0.25}$HfS$_2$ (optimized lattice parameters):}\\
Space group:	 $P\bar3m1$ (164)											&&Hf	&&0		&0		&0	\\
$k$-mesh for structure optimization: $14\times14\times11$						&&Hf	&&0		&-1/2	&0	\\
$k$-mesh for band structure/DOS/Optics: $29\times29\times23$					&&S		&&0.167	&-0.167	&0.190	\\
Lattice parameters: $a=b=\SI{7.08}{\angstrom},\,c=\SI{7.80}{\angstrom}$	&&S		&&-1/3	&1/3	&0.189	\\
																		&&K		&&0		&0		&-1/2	\\									
\\
\underline{5. Bulk K$_{0.33}$HfS$_2$ (optimized lattice parameters):}\\
Space group:	 $P\bar31m$ (162)											&&Hf	&&0		&0		&0	\\
$k$-mesh for structure optimization: $15\times15\times11$						&&Hf	&&1/3	&-1/3	&0	\\
$k$-mesh for band structure/DOS/Optics: $33\times33\times23$					&&S		&&0		&-0.334	&0.191	\\
Lattice parameters: $a=b=\SI{6.13}{\angstrom},\,c=\SI{7.79}{\angstrom}$	&&K		&&0		&0		&-1/2	\\
\\
\underline{6. Bulk K$_{0.50}$HfS$_2$ (optimized lattice parameters):}\\
Space group: $P1\,2/m1$ (10)												&&Hf	&&0		&0		&0	\\
$k$-mesh for structure optimization: $11\times24\times13$						&&Hf	&&0		&1/2	&-1/2	\\
$k$-mesh for band structure/DOS/Optics: $25\times18\times27$					&&S		&&-0.196	&0	&-0.337	\\
Lattice parameters: $a=\SI{6.15}{\angstrom},\,b=\SI{3.55}{\angstrom},\,c=\SI{7.66}{\angstrom}$	&&S	&&0.196	&1/2	&-0.164	\\
Axis	angles: $\alpha=\beta=\gamma=\SI{90}{\degree}$						&&K		&&-1/2		&0		&0	\\
\\
\underline{7. Bulk K$_{0.66}$HfS$_2$ (optimized lattice parameters):}\\
Space group: $P\bar31m$ (162)											&&Hf	&&0		&0		&0	\\
$k$-mesh for structure optimization: $15\times15\times11$						&&Hf	&&1/3	&-1/3	&0	\\
$k$-mesh for band structure/DOS/Optics: $32\times32\times22$					&&S		&&0		&-0.332	&0.198	\\
Lattice parameters: $a=b=\SI{6.15}{\angstrom},\,c=\SI{7.61}{\angstrom}$	&&K		&&-1/3	&1/3	&-1/2	\\
\\
\underline{8. Bulk K$_{0.75}$HfS$_2$ (optimized lattice parameters):}\\
Space group: $P\bar3m1$ (164)		 									&&Hf	&&0		&0		&0	\\
$k$-mesh for structure optimization: $14\times14\times12$	 					&&Hf	&&0		&-1/2	&0	\\
$k$-mesh for band structure/DOS/Optics: $28\times28\times23$					&&S		&&0.166	&-0.166	&0.198	\\
Lattice parameters: $a=b=\SI{7.13}{\angstrom},\,c=\SI{7.57}{\angstrom}$	&&S		&&-1/3	&1/3	&0.200	\\
																		&&K		&&-1/2	&0		&-1/2	\\
\\
\pagebreak
\underline{9. Bulk K$_{0.83}$HfS$_2$ (optimized lattice parameters):}\\
Space group: $P\bar31m$ (162)		 									&&Hf		&&0		&0 		&0	\\
$k$-mesh for structure optimization: $10\times10\times14$	 					&&Hf		&&-0.330 &-0.165 &0	\\
Lattice parameters: $a=b=\SI{12.37}{\angstrom},\,c=\SI{7.56}{\angstrom}$	&&Hf		&&1/2 	&1/2 	&0	\\
																		&&Hf		&&-1/3	&1/3 	&0	\\
																	&&S		&&-0.500 &0.333 &0.196	\\
																	&&S		&&0 	&0.333 &0.201	\\
																	&&S		&&0 &-0.167 &0.204	\\
																	&&K		&&0.349 &0.175 &-1/2	\\
																	&&K		&&1/2		&1/2		&-1/2	\\
																	&&K		&&0		&0		&-1/2	\\
\\
\underline{10. Bulk K$_{1.00}$HfS$_2$ (optimized lattice parameters):}\\
Space group: $P\bar3m1$ (164)										&&Hf	&&0		&0		&0	\\
$k$-mesh for structure optimization: $24\times24\times10$					&&S		&&1/3	&-1/3	&0.202	\\
$k$-mesh for band structure/DOS/Optics: $52\times52\times22$				&&K		&&0		&0		&1/2	\\
Lattice parameters: $a=b=\SI{3.59}{\angstrom},\,c=\SI{7.50}{\angstrom}$	\\
\\
\underline{11. Surface HfS$_2$ (5 structural unit cells and $\SI{20}{\angstrom}$} \\
\underline{vacuum):}\\
space group: $P\bar3m1$ (164)		 								&&Hf		&&0		&0		&0.250	\\
$k$-mesh for band structure/DOS/Optics: $52\times52\times3$		 		&&Hf		&&0 	&0 		&0.375	\\
Lattice parameters: $a=b=\SI{3.56}{\angstrom},\,c=\SI{45.76}{\angstrom}$						&&Hf		&&0 	&0 	&1/2	\\
																	&&S		&&1/3&-1/3&0.283	\\
																	&&S		&&1/3&-1/3&0.407	\\
																	&&S		&&1/3&-1/3&-0.468	\\
																	&&S		&&1/3&-1/3&-0.343	\\
																	&&S		&&1/3&-1/3&-0.219	\\
\end{longtable*} 

\setlength\LTcapwidth{\textwidth}
\begin{longtable*}{l>{\centering\arraybackslash}p{0.1cm}c>{\centering\arraybackslash}p{0.1cm}>{\centering\arraybackslash}p{2.6cm}>{\centering\arraybackslash}p{2.6cm}>{\centering\arraybackslash}p{2.6cm}}
\caption{Optimized atomic coordinates, experimental/optimized lattice parameters, space groups and $k$-meshes for K$_x$HfSe$_2$. The $k$-meshes were chosen such that the $k$-point density is approximately the same in all structures.}
\label{tab:HfSe2_CalcDetails}\\
\hline\hline
					&&&\multicolumn{4}{c}{Optimized atomic coordinates}\\ \cline{5-7}
Calculation parameters	&&Atom 	&&x/a		&y/b	&z/c\\ 
\hline
\endhead
\hline\hline
\endfoot
\\
\underline{1. Bulk HfSe$_2$ (experimental lattice parameters):}	\\
Space group: $P\bar3m1$ (164)											&&Hf	&&0		&0		&0	\\
$k$-mesh for structure optimization: $24\times24\times13$						&&Se	&&1/3	&-1/3	&0.255	\\
$k$-mesh for band structure/DOS/Optics: $52\times52\times28$	\\
Lattice parameters: $a=b=\SI{3.76}{\angstrom},\,c=\SI{6.06}{\angstrom}$ \\
\\
\underline{2. Bulk HfSe$_2$ (optimized lattice parameters):}					\\ 
Space group: $P\bar3m1$ (164)											&&Hf	&&0		&0		&0	\\
$k$-mesh for structure optimization: $24\times24\times13$						&&Se	&&1/3	&-1/3	&0.262	\\
$k$-mesh for band structure/DOS/Optics: $52\times52\times28$		\\
Lattice parameters: $a=b=\SI{3.68}{\angstrom},\,c=\SI{6.02}{\angstrom}$ \\ 
\\
\underline{3. Bulk K$_{0.17}$HfSe$_2$ (optimized lattice parameters):}\\
Space group:	 $P1\,2/m1$ (10)												&&Hf		&&0		&0		&0	\\
	$k$-mesh for structure optimization: $12\times9\times14$					&&Hf		&&0		&-1/2		&-1/2	\\
	$k$-mesh for band structure/DOS/Optics: $25\times18\times27$				&&Hf		&&0		&-0.337		&0	\\
Lattice parameters: $a=\SI{6.36}{\angstrom},\,b=\SI{11.01}{\angstrom},\,c=\SI{7.89}{\angstrom}$	&&Hf	&&0		&0.171		&-1/2	\\
Axis	angles: $\alpha=\beta=\gamma=\SI{90}{\degree}$						&&Se	&&-0.201	&-0.168	&-0.167	\\
																		&&Se	&&0.203 	&-0.334 &-0.332	\\
																		&&Se	&&-0.204 	&-1/2 			&-0.168	\\
																		&&Se	&&-0.199 	&0				&0.337	\\
																		&&K	&&1/2		&0		&0	\\
\\ 
\underline{4. Bulk K$_{0.25}$HfSe$_2$ (optimized lattice parameters):}\\
Space group:	 $P\bar3m1$ (164)											&&Hf	&&0		&0		&0	\\
$k$-mesh for structure optimization: $14\times14\times11$						&&Hf	&&0		&-1/2		&0	\\
$k$-mesh for band structure/DOS/Optics: $28\times28\times23$					&&Se	&&0.167	&-0.167	&0.196	\\
Lattice parameters: $a=b=\SI{7.33}{\angstrom},\,c=\SI{8.12}{\angstrom}$	&&Se	&&-1/3	&1/3	&0.195	\\
																		&&K		&&0		&0		&-1/2	\\									
\\
\underline{5. Bulk K$_{0.33}$HfSe$_2$ (optimized lattice parameters):}\\
Space group:	 $P\bar31m$ (162)											&&Hf	&&0		&0		&0	\\
$k$-mesh for structure optimization: $15\times15\times11$						&&Hf	&&1/3		&-1/3		&0	\\
$k$-mesh for band structure/DOS/Optics: $32\times32\times22$					&&Se	&&0	&-0.335	&0.197\\
Lattice parameters: $a=b=\SI{6.35}{\angstrom},\,c=\SI{8.12}{\angstrom}$	&&K		&&0		&0		&1/2	\\
\\
\underline{6. Bulk K$_{0.50}$HfSe$_2$ (optimized lattice parameters):}\\
Space group: $P1\,2/m1$ (10)												&&Hf	&&0		&0		&0	\\
$k$-mesh for structure optimization: $11\times24\times13$						&&Hf	&&0		&1/2		&-1/2	\\
$k$-mesh for band structure/DOS/Optics: $24\times52\times26$					&&Se	&&-0.199	&0	&-0.335	\\
Lattice parameters: $a=\SI{6.37}{\angstrom},\,b=\SI{3.68}{\angstrom},\,c=\SI{7.99}{\angstrom}$	&&Se&&0.201 &1/2 &-0.168	\\
Axis	angles: $\alpha=\beta=\gamma=\SI{90}{\degree}$						&&K		&&-1/2		&0		&0	\\
\\
\underline{7. Bulk K$_{0.66}$HfSe$_2$ (optimized lattice parameters):}\\
Space group: $P\bar31m$ (162)											&&Hf	&&0		&0		&0	\\
$k$-mesh for structure optimization: $15\times15\times11$						&&Hf	&&1/3	&-1/3		&0	\\
$k$-mesh for band structure/DOS/Optics: $31\times31\times22$					&&Se	&&0		&-0.332	&0.199	\\
Lattice parameters: $a=b=\SI{6.40}{\angstrom},\,c=\SI{7.93}{\angstrom}$	&&K		&&-1/3	&1/3		&1/2	\\
\\
\underline{8. Bulk K$_{0.75}$HfSe$_2$ (optimized lattice parameters):}\\
Space group: $P\bar3m1$ (164)		 									&&Hf	&&0		&0		&0	\\
$k$-mesh for structure optimization: $13\times13\times10$	 					&&Hf	&&0		&-1/2		&0	\\
$k$-mesh for band structure/DOS/Optics: $27\times27\times22$					&&Se	&&0.166	&-0.166	&0.202	\\
Lattice parameters: $a=b=\SI{7.41}{\angstrom},\,c=\SI{7.89}{\angstrom}$	&&Se	&&-1/3	&1/3		&0.204	\\
																		&&K		&&-1/2		&0		&-1/2	\\
\\
\underline{9. Bulk K$_{0.83}$HfSe$_2$ (optimized lattice parameters):}\\
Space group: $P\bar31m$ (162)		 									&&Hf	&&0	&0 	&0	\\
$k$-mesh for structure optimization: $7\times7\times10$	 					&&Hf	&&-0.330 &-0.165 &0	\\
Lattice parameters: $a=b=\SI{12.87}{\angstrom},\,c=\SI{7.88}{\angstrom}$	&&Hf	&&-1/2 	&-1/2 	&0	\\
																		&&Hf	&&-1/3	&1/3 	&0	\\
																		&&Se	&&-0.499 &0.333 &0.200	\\
																		&&Se	&&0 &0.333 &0.204	\\
																		&&Se	&&0 &-0.167 &0.207	\\
																		&&K		&&0.348 &0.174 &1/2	\\
																		&&K		&&-1/2		&-1/2		&1/2	\\
																		&&K		&&0		&0		&1/2	\\
\\
\underline{10. Bulk K$_{1.00}$HfSe$_2$ (optimized lattice parameters):}\\
Space group: $P\bar3m1$ (164)											&&Hf	&&0		&0		&0	\\
$k$-mesh for structure optimization: $24\times24\times10$						&&Se	&&1/3	&-1/3	&0.203	\\
$k$-mesh for band structure/DOS/Optics: $51\times51\times22$					&&K		&&0		&0		&-1/2	\\
Lattice parameters: $a=b=\SI{3.74}{\angstrom},\,c=\SI{7.82}{\angstrom}$	\\
\\
\pagebreak
\underline{11. Surface HfSe$_2$ (5 structural unit cells and $\SI{20}{\angstrom}$} \\
\underline{vacuum):}\\
Space group: $P\bar3m1$ (164)		 									&&Hf		&&0		&0		&0.245	\\
$k$-mesh for band structure/DOS/Optics: $52\times52\times3$		 			&&Hf		&&0 	&0 		&0.373	\\
Lattice parameters: $a=b=\SI{3.68}{\angstrom},\,c=\SI{47.24}{\angstrom}$	&&Hf		&&0 	&0 	&-1/2	\\
																		&&Se	&&1/3&-1/3&0.279	\\
																		&&Se	&&1/3&-1/3&0.406	\\
																		&&Se	&&1/3&-1/3&-0.467	\\
																		&&Se	&&1/3&-1/3&-0.339	\\
																		&&Se	&&1/3&-1/3&-0.212	\\
\end{longtable*} 

\begin{table*}
\centering
\renewcommand{\tabcolsep}{0.2cm}
\caption{Measured momentum transfer values for diffraction peak positions with Miller indices [11$w$] and the unit cell parameters in the $c$-direction calculated from those values.}
\label{tab:DiffrPP}
\begin{threeparttable}
\begin{tabular}{cccccccccccccc}
\toprule
\toprule
\multicolumn{2}{c}{}	& \multicolumn{5}{c}{K$_x$HfS$_2$} &&&\multicolumn{5}{c}{K$_x$HfSe$_2$}\\ \cline{3-7} \cline{10-14}

\multicolumn{2}{c}{}	& \multicolumn{2}{c}{Undoped ($x=0.00$)} &\multicolumn{1}{c}{} &\multicolumn{2}{c}{Doped ($x=0.55$)} &&\multicolumn{1}{c}{}	& \multicolumn{2}{c}{Undoped ($x=0.00$)} &\multicolumn{1}{c}{} &\multicolumn{2}{c}{Doped ($x=0.70$)} 
\\ \cline{3-4} \cline{6-7} \cline{10-11} \cline{13-14}

\multicolumn{1}{c}{Miller}	&&	\multicolumn{1}{c}{Momen-}	& \multicolumn{1}{c}{Calculated} &\multicolumn{1}{c}{} &	\multicolumn{1}{c}{Momen-}	& \multicolumn{1}{c}{Calculated} && \multicolumn{1}{c}{}	&	\multicolumn{1}{c}{Momen-}	& \multicolumn{1}{c}{Calculated} &\multicolumn{1}{c}{} &	\multicolumn{1}{c}{Momen-}	& \multicolumn{1}{c}{Calculated}\\

\multicolumn{1}{c}{Index}	&&	\multicolumn{1}{c}{tum}	& \multicolumn{1}{c}{Unit Cell} &\multicolumn{1}{c}{} &	\multicolumn{1}{c}{tum}	& \multicolumn{1}{c}{Unit Cell} && \multicolumn{1}{c}{}	&	\multicolumn{1}{c}{tum}	& \multicolumn{1}{c}{Unit Cell} &\multicolumn{1}{c}{} &	\multicolumn{1}{c}{tum}	& \multicolumn{1}{c}{Unit Cell}\\

\multicolumn{1}{c}{$w$\hspace{0.5pt}-value}	&&	\multicolumn{1}{c}{Transfer}	& \multicolumn{1}{c}{Parameter} &\multicolumn{1}{c}{} &	\multicolumn{1}{c}{Transfer}	& \multicolumn{1}{c}{Parameter} &&&	\multicolumn{1}{c}{Transfer}	& \multicolumn{1}{c}{Parameter} &\multicolumn{1}{c}{} &	\multicolumn{1}{c}{Transfer}	& \multicolumn{1}{c}{Parameter}\\

\multicolumn{1}{c}{([11$w$])} &&	 \multicolumn{1}{c}{$|\mathbf{q}|\:[\SI{}{\per\angstrom}]$}	& \multicolumn{1}{c}{ c\tnote{a}\enspace$[\SI{}{\angstrom}]$} &\multicolumn{1}{c}{} &	 \multicolumn{1}{c}{$|\mathbf{q}|\:[\SI{}{\per\angstrom}]$}	& \multicolumn{1}{c}{ c\tnote{a}\enspace$[\SI{}{\angstrom}]$} &&&	 \multicolumn{1}{c}{$|\mathbf{q}|\:[\SI{}{\per\angstrom}]$}	& \multicolumn{1}{c}{c\tnote{a}\enspace$[\SI{}{\angstrom}]$} &\multicolumn{1}{c}{} &	 \multicolumn{1}{c}{$|\mathbf{q}|\:[\SI{}{\per\angstrom}]$}	& \multicolumn{1}{c}{ c\tnote{a}\enspace$[\SI{}{\angstrom}]$}\\
\midrule
0&&3.47&- & &3.49&-&&&3.35&- & &3.33&-\\
1&&3.63&5.76& &3.58&7.93&&&3.51&5.93& &3.41&8.26\\
2&&4.09&5.77& &3.85&7.75&&&3.92&6.14& &3.66&8.19\\
3&&4.76&5.77& &4.24&7.85&&&4.55&6.12& &4.05&8.18\\
4&&- &- & &4.75&7.81&&&- &- & &4.50&8.31\\
\midrule
\multicolumn{2}{l}{Average}	&&5.77	&				&&7.84	&&&	&6.06	&				&&8.23	\\
\bottomrule
\bottomrule
\end{tabular}
	\begin{tablenotes}
		\item [a] Calculation of unit cell parameter: $c=2 \pi w/\sqrt{q_{[110]}^2-q_{[11w]}^2}$ where $q_{[11w]}$ is the momentum transfer associated with the diffraction peak of the Miller index [11$w$].
\end{tablenotes}
\end{threeparttable}
\end{table*}
\begin{figure*}[!]
	\includegraphics [width=0.9\textwidth]{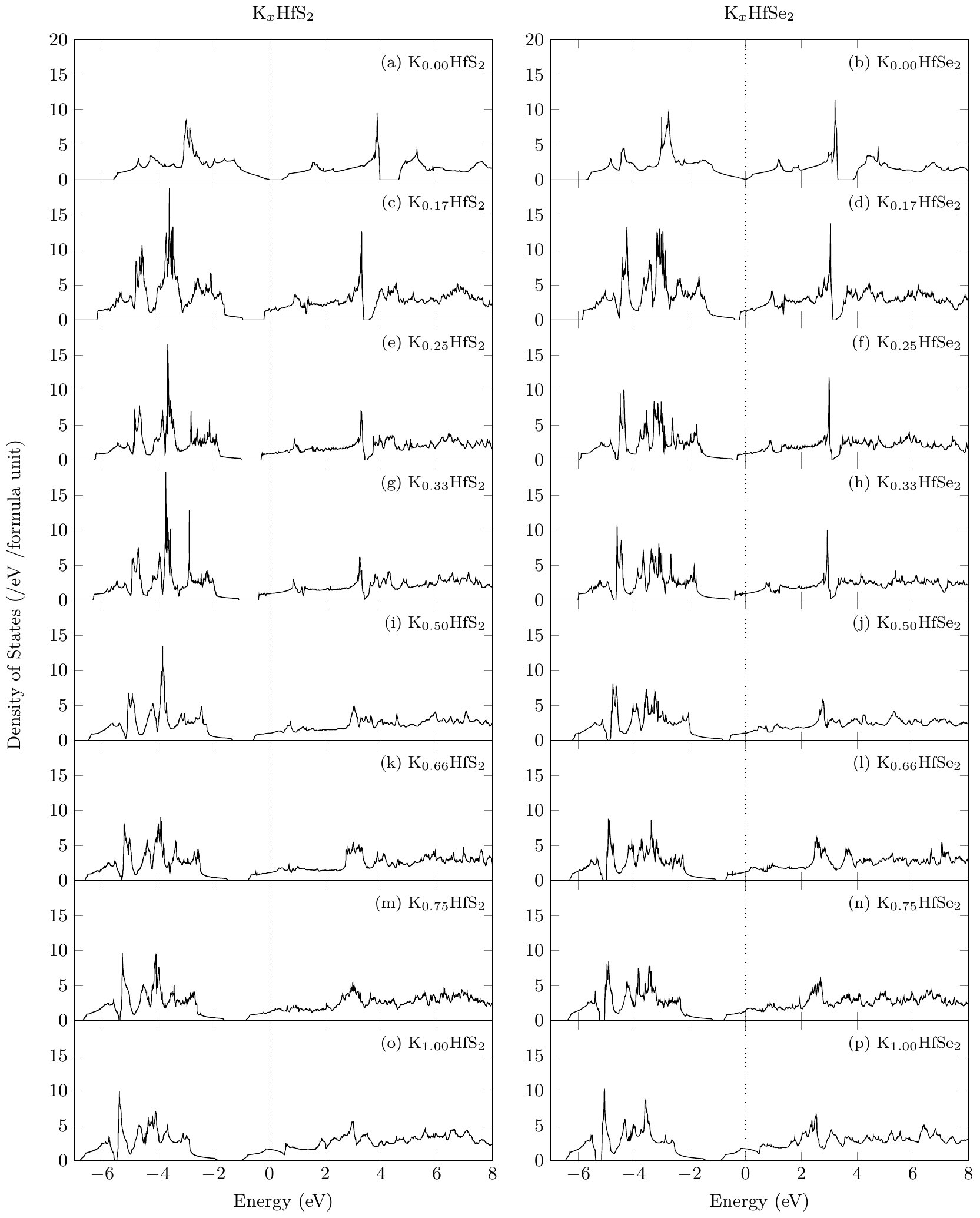}
	\caption{(Color online) Density of states for K$_x$HfS$_2$ and K$_x$HfSe$_2$ based on optimized lattice parameters.}
	\label{fig:DOS-All}
\end{figure*}
%
%
\begin{table*}
\centering
\renewcommand{\tabcolsep}{0.265cm}
\caption{Drude-Lorentz model fit parameters for the optical conductivity function of K$_{0.55}$HfS$_2$ and K$_{0.70}$HfSe$_2$. 
}
\label{tab:FitParameters}
\begin{tabular}{ccrrrccrrrr}
\toprule
\toprule
Oscillator &\multicolumn{1}{c}{Oscillator}	&	&\multicolumn{3}{c}{K$_{0.55}$HfS$_2$}&&\multicolumn{3}{c}{K$_{0.70}$HfSe$_2$}\\ \cline{4-6} \cline{8-10}

index $i$	&\multicolumn{1}{c}{Type}	&&	\multicolumn{1}{c}{$\omega_i$ (eV)} &	\multicolumn{1}{c}{$\gamma_i$ (eV)}	&	\multicolumn{1}{c}{$\omega_{pi}$ (eV)} &	&	\multicolumn{1}{c}{$\omega_i$ (eV)} &	\multicolumn{1}{c}{$\gamma_i$ (eV)}	&	\multicolumn{1}{c}{$\omega_{pi}$ (eV)}\\
\midrule
-&Drude &&0&0.37&3.55&&0&0.25&3.89\\
1&Lorentz &&3.14&0.33&1.61&&2.2&0.14&0.56\\
2&Lorentz &&3.34&0.36&1.95&&2.37&0.23&1.11\\
3&Lorentz &&3.6&0.46&1.69&&2.53&0.28&1.95\\
4&Lorentz &&3.95&0.98&2.38&&2.68&0.24&1.51\\
5&Lorentz &&5.32&1.45&2.87&&2.85&0.43&1.99\\
6&Lorentz &&6.2&0.8&2.32&&3.19&0.69&2.28\\
7&Lorentz &&6.68&0.19&0.61&&3.66&0.49&0.87\\
8&Lorentz &&6.95&1.78&6.61&&4.42&1.3&2.81\\
9&Lorentz &&7.55&0.31&0.96&&5.81&1.98&7.28\\
10&Lorentz &&7.85&0.17&0.68&&7.39&3.71&9.71\\
11&Lorentz &&8.34&2.84&7.87&&9.75&1.64&2.68\\
12&Lorentz &&10.26&3.33&5.11&&12.83&7.76&9.66\\
13&Lorentz &&14.14&5.31&6.12&&19.93&4.94&3.87\\
14&Lorentz &&20.74&23.72&13.05&&25.64&10.29&8.66\\
\bottomrule
\bottomrule
\end{tabular}
\end{table*}
%


\end{document}